\documentclass[12pt,a4paper,dvips]{article}
\usepackage{a4p}
\usepackage{cite,mcite}
\usepackage{graphicx,amsmath}
\usepackage{pennames}
\usepackage{physics }
\usepackage{l3_titlePH,ifthen}
\usepackage{endfig}
\def\Pap{\overline{p}}
\def\Pp{p}
\journalname{Astroparticle Physics}
\date{December 8, 2004}

\preprint{2004-076}
\graphicspath{{/l3/paper/example/figs/}
              {/afs/cern.ch/user/p/parriaud/public/tex/publication/}}
\newlength{\capindent}
\newlength{\capwidth}
\newlength{\figwidth}
\newcommand{\icaption}[2][!*!,!]{\hspace*{\capindent}%
  \begin{minipage}{\capwidth}
    \ifthenelse{\equal{#1}{!*!,!}}%
      {\caption{#2}}%
      {\caption[#1]{#2}}
  \end{minipage}}
\begin{document}
\begin{titlepage}
\title{Measurement of the Shadowing of High-Energy Cosmic Rays by the Moon: A Search for TeV-Energy Antiprotons}

\author{The L3 Collaboration}



%
%
\begin{abstract}
 
The shadowing of high-energy cosmic rays by the Moon has been observed with a significance of 9.4 standard deviations 
with the L3+C muon spectrometer at CERN. A significant effect of the Earth magnetic field is observed. 
Since no event deficit on the east side of the Moon has been observed, an upper limit at 90\,\% confidence level on the 
antiproton to proton ratio of 0.11 is obtained for primary energies around 1~TeV.

\end{abstract}
%



%
\submitted
\end{titlepage}
%
%

\section{Introduction}

\paragraph{Physics motivation}

The effect of the Moon, or the Sun, on cosmic rays was first noted
by Clark in 1957 \cite{clark}. As these bodies pass overhead
they block the particles, so their shadows in the cosmic ray flux
must be visible by detectors on Earth. \\
However the first observation
of such a shadowing had to wait for the results of the CYGNUS collaboration
in 1991 \cite{cygnus1}. There are two reasons for this long delay.
First, the particles must be insensitive or weakly sensitive
to the Earth magnetic field. Obvious candidates are $\gamma$-rays or
energetic cosmic ray particles.
The former are very rare and the observation of the latter 
above the nearly-isotropic large background of low-energy cosmic rays, was only
possible with the advent of large Extended-Air-Shower (EAS) detectors,
able to collect high statistics.
Second, a crucial parameter is the angular resolution of the detectors. The signal over background
ratio is inversely proportional to the square of this angular resolution and
events are spread out from the expected position
due to the finite angular resolution. The performance
of the detector has to cope with the angular radius of the Moon (or
the Sun), each having approximately a $0.27^{\circ}$ radius, and
only at the beginning of the 90's, the angular
resolutions of cosmic ray detectors reached the one-degree level.\\
Since then, several other experiments, both EAS arrays and large
underground detectors have been able to see the Moon-shadow effect
\mcite{casa1,tibet1,macro1,macro2,soudan1}. The observation is used for a check
of the angular resolution of the apparatus and, by comparing the
observed position of the deficit to the expected Moon position, to
evaluate systematic pointing errors. The understanding of the
alignment and of the angular resolution is a key issue for any 
point-source search.

In 1990, a more challenging use of the Moon shadow effect has been
proposed \cite{urban}. The use of the Moon collimation, together
with the Earth magnetic field, allows a charge determination.
Negatively charged primaries are deflected towards the west and 
positively charged primaries towards the east. 
If antiprotons are present in the cosmic ray flux,
they will generate a shadow on the opposite side of the
Moon relative to the shadow from cosmic rays induced from matter.
This article discusses a search for cosmic ray antiprotons
using the L3+C muon spectrometer of
the L3 detector at the CERN LEP accelerator.
Published direct measurements of the $\rm \Pap/\Pp$ ratio
exist only below $40\rm\;GeV$ and this method is sensitive at TeV energies.
Only non-standard sources would be the origin of such high-energy antiprotons.
No such study has yet been
published using the Earth-Moon system as a spectrometer: EAS arrays 
(with the exception of the Tibet array) and
underground detectors have a too-high detection threshold, so that the
effect of the Earth magnetic field is just a small perturbation. This
is no longer the case in the L3+C experiment.
Due to only 30 m of overburden,
multiple scattering remains small even for low energies and
the accumulation rate is much larger than in other underground detectors.
Moreover, the measurement of the muon
momentum allows for an off-line tuning of the threshold,
leading to a possible optimisation of the shadow effect.
The present study exploits these possibilities and looks at the conditions 
to set a limit to the $\rm \Pap$ content in the TeV region.

This section describes the status of cosmic ray antiproton data and summarizes    
other experimental observations of the Moon shadow by cosmic rays.
The experimental setup is presented in Section 2.
Section 3 contains a review of the main parameters involved in a Moon-shadow
experiment and describes the role of the different
Monte Carlo simulations. In particular, the simulation of the
experimental angular-resolution is checked with the help of
two-track events. Data and backgrounds are presented in Section 4.
The observation and interpretation of the deficit of events in the
Moon direction are described in Section 5. Results concerning
the experimental angular-resolution and a possible shadowing effect
due to antiprotons are discussed. Conclusions are
presented in Section 6.

\paragraph{Cosmic-ray antiprotons}

The experimental $\rm \Pap/\Pp$ ratio below 50 GeV
is compatible with a secondary origin of the
cosmic ray antiprotons \cite{moska}. Data are obtained by balloon-borne experiments
and, recently, satellite experiments. The CAPRICE \cite{caprice1} and HEAT
\cite{heat1} collaborations obtained ratios measured  
between 4 and 50 GeV with a series of balloon flights.
The CAPRICE data
do not show any flattening of the ratio
with increasing energy (ratio $\sim 10^{-3}$ around $40\rm\;GeV$) as expected 
from the secondary production model,
in contrast with the last HEAT data point, which sets a limit of $ \sim 2 \cdot 10^{-4}$ above $20\rm\;GeV$.
This underlines the limited statistics
available up to now in this kind of experiments.

The uncertainties of theoretical models are large below the well defined flux maximum
of $2\rm\;GeV$ due to the complexity of production and propagation at low energy.
On the contrary, the secondary high-energy flux is predicted with good confidence,
all estimates being consistent with each other. Above a few tens of GeV,
the antiproton production becomes quite negligible, the flux falling by 3
orders of magnitude below the maximum for antiproton energies around 40 GeV.
Any experimental hint of antiprotons in these high energy regions
would therefore be of prime importance.

First upper limits \cite{ISVHECRI02} on the $\rm \Pap/\Pp$ ratio around 1 TeV were presented by the  L3+C 
and the TIBET-AS$\gamma$ - collaborations \cite{TibetRatio} at different 
conferences. The Tibet array has a worse angular resolution compared to the L3+C experiment and observes a smaller 
deviation of the Moon shadow due to its sensitivity to higher primary energies.

In reference \cite{steph} an upper limit on the $\rm \Pap/\Pp$ ratio
around 1 TeV is calculated from different measured $\mu^{+}/\mu^{-}$ - ratios at ground level
with large uncertainties.
This indirect determination of a limit is based on cascade calculations which depend
on the assumed primary composition and the hadronic interaction cross-sections at high energies 
\cite{ECRS}. The method has therefore large systematic uncertainties.
L3+C has recently measured precisely the muon momentum spectrum, 
as well as the charge ratio and the angular dependence \cite{L3+Cspec}. 
Based on to-day's knowledge of the parameters entering the calculation, an estimate
of an upper limit on the  contribution of antiprotons to the primary flux 
could not compete with the one presented in this paper. 

There are at least 3 models of exotic sources able to produce high
energy antiprotons: primordial black hole (PBH) evaporation \cite{sala},
dark-matter neutralino annihilation \cite{ullio1,ullio2} and
high-energy antiprotons from extragalactic sources \cite{steph,stecker}.
In some scenarios, the last two models can provide a $\rm \Pap/\Pp$ ratio increase
up to the 10\,\% level in the energy range under investigation. 
This shows the importance of measuring the antiproton flux at high
energy. The antiproton energy-spectrum
is of course expected to be different for each type of sources.
Putting a limit to the number of antiprotons will, in any case, constrain
some of the parameters of the models. \\

\paragraph{Moon shadow experiments}

EAS arrays were the first detectors to
look for a Moon-shadow effect and the first observation was reported by
the CYGNUS collaboration \cite{cygnus1}, with  
a 4.9 standard deviations $(s.d.)$ significance.
The CASA group\cite{casa1}, with a larger array, obtained about the same result.
In absence of any dominant point-source, the
Moon-shadow experiment provides a unique possibility to make a direct
measurement of the angular resolution of the detectors and to verify the 
pointing accuracy.

Large underground detectors also have the potential to
observe the Moon-shadow.
Backtracking  muons to the surface depends on the correct estimation of
multiple Coulomb scattering in the large rock overburden,
a process known to be essentially non-Gaussian.
Results have been presented
by the MACRO \mcite{macro1,macro2} and
SOUDAN \cite{soudan1} collaborations.
These detectors have muon energy-thresholds of several TeV.
As a consequence, statistics are low even if the data were accumulated for
nearly 10 years. 
The effect of the geomagnetic field is hardly
visible being less important than the observed shift from the origin
due to pointing uncertainties.
To make
the influence of the deflection due to the magnetic field significant,
the only possibility is to lower the detection threshold of the
primary particles. 
Locating EAS arrays at very high altitudes is a
solution. The TIBET air-shower experiment \cite{tibet1} has been
operated since 1990 at 4300 m above sea level ({\it a.s.l.}) and has provided 
the first unambiguous effect of the geomagnetic field on the Moon shadow.

Imaging-Cherenkov detectors have also been proposed for the
observation of the Moon shadow \cite{urban}.
The search of $\gamma$ sources with this technique was a success,
in particular with the observation of
the Crab nebula and a handful of other point sources.
The application to the Moon-shadow measurement is
more difficult as moonlight prohibits the use of visible photons and
no Moon shadow was observed with this technique
\cite{artemis}.

Another promising technique uses a large volume of
water as the detection medium. Photomultiplier tubes 
detect the Cherenkov radiation produced in the water by
relativistic charged particles or photons produced in the primary
shower. The MILAGRO collaboration
\mcite{milagro_1,milagro_2} built a first
prototype, MILAGRITO, running from February 1997 to May 1998,
then a full detector, MILAGRO, starting its operation in
February 1999. The goal is to be sensitive to primary cosmic rays down to
$1\rm\;TeV$ or less, as imaging-Cherenkov detectors, while maintaining an all-sky acceptance
 and a high-duty cycle like EAS arrays.
Preliminary results  have shown that the Moon shadow effect is
observed with a significance  above 20~$s.d.$  \cite{milagro_pre}.
No result of an antiproton search is yet available.

%
%
\section{The L3+C detector}

The L3+C detector is part of the L3 apparatus \cite{l3}, one of the four particle
detectors  installed on the LEP Collider. It is located
underneath the French-Swiss border at CERN, at 450 m $a.s.l$.
under $30\rm\;m$ of sedimentary rocks called molasse
(density $\sim 7.2 \times 10^{3}\rm\;g/cm^{2}$).
It mainly makes use of the
muon chamber system which was designed to make a very precise measurement
of muons produced in $\rm e^{+}e^{-}$ collisions.
The muon spectrometer consists of two octagonally shaped rings, each with
eight ``octants'', installed in a 12 m diameter solenoidal magnet which
provides a uniform field of 0.5 Tesla along the $\rm e^{+}e^{-}$ beam direction.
Each octant contains precision drift chambers organised in three layers to
measure the projection of the muon trajectory onto the plane orthogonal 
to the magnetic field, and layers of drift cells to measure the projection
along the magnetic field direction.
Other parts of the L3 detector are not used by L3+C.
To fulfil the specific features of the cosmic ray experiment and to
make the running of both L3 and L3+C completely independent
from each other, several systems are specifically
added to the L3 setup:
\begin{itemize}
 \item 
     On top of the magnet, $\rm 202~ m^2$ of plastic scintillators
     are installed to determine the muon arrival time.
 \item
      A new trigger and data-acquisition system is built to decouple the
      L3 and L3+C operation.
 \item 
      A precise timing system is devised. It is based on an external
      GPS module and it includes also 1Hz and 10 MHz clocks. 
 \item
      An air-shower scintillator array is installed on the roof of the
      surface building to estimate the shower size associated to a
      detected muon. Its data are not used in the present analysis.
\end{itemize}
The geometrical acceptance of the detector amounts to about
$200\rm\;m^{2}sr$ and the muon momentum threshold set by the overburden is $15\rm\;GeV$.
The detector was operational in May 1999 and a total of $1.2 \times 10^{10}$
muon triggers were collected up to November 2000, corresponding to
an effective live-time of 312 days.
Both momentum resolution
and detection efficiency are checked using muons from Z decays
that L3+C could also detect when the accelerator was running at a
centre-of-mass energy equal to the Z-boson mass. Studies are extended
using muons going through two separate octants, both giving a nearly
independent measurement of the particle momentum.
The momentum resolution is found to be 4.6\,\% at $45\rm\;GeV$ and
7.4\,\% at $100\rm\;GeV$.
The muon momentum threshold can be adjusted off-line to optimise the results.
A detailed description of the L3+C detector and its performances
is given in References \cite{nim,L3+Cspec}.

\section{Experimental considerations and Monte Carlo simulations}

A specific description \cite{Moonshadow} of the Earth-Moon spectrometer system
and of the cosmic ray shadowing effect is implemented in the
L3+C simulation with two aims:
\begin{itemize}
\item to take into account as accurately as possible the different
detector components, the Earth magnetic field,
the cosmic ray showering in the atmosphere, the multiple scattering of muons in the molasse,
and the reconstruction, 
\item to understand the relative importance of the various parameters
contributing to the measurement of the $\rm \Pap/\Pp$ ratio.
\end{itemize}

The detector properties are described with a Monte Carlo 
based on the GEANT program \cite{geant} which takes into account the
effect of energy loss, multiple Coulomb scattering and showering in the
detector.
The basic version of the model is identical to the L3 simulation
package, but specific features required by the L3+C setup, 
are taken into account, such as the additional scintillators on top
of the magnet, or the magnetic field in the coil and yoke of the
magnet. For the measurement of cosmic rays originating from the
atmosphere, the overburden above the detector
must also be included. The energy loss of muons and the
smearing of their angular direction is an important issue. The
whole surrounding of the L3 detector, consisting mainly of molasse,
is introduced, including the access shafts to the experimental cavern and the shielding structures.
All main physics processes related to the muon propagation through matter, such as multiple scattering,
secondary-particle production including $\delta$-rays, pair
production, energy loss and decay are fully simulated.

Special attention is
put on the simulation of the muon chambers, by including all
inefficiencies due to less efficient and dead cells in the muon detector.
The simulated Monte Carlo events are reconstructed and backtracked
to the ground level in the same way as the data events.

The simulation of the detector is based on the generation of muons.
Instead of performing a full simulation of the air-shower cascade
generated by the primary cosmic radiation in the atmosphere, single and
double muons are generated above the detector, according to the known
angular and energy distributions obtained by a full air shower
simulation using the CORSIKA package \cite{corsika}. The interactions,
decays, annihilations and secondary-particle production in the air
are fully simulated, according to the current experimental
knowledge and to various theoretical models. 

 \subsection{Angular resolution}

One of the key issues of the Moon-shadow measurement is the experimental angular resolution. The Moon
subtends a radius of $0.27^{\circ}$ and the angular resolution has to match this
constraint. The following
factors are taken into consideration:

\begin{itemize}
  \item the muon direction with respect to the primary nucleon direction,
  \item the multiple scattering in the molasse above the detector,
  \item the intrinsic
angular resolution due to the muon chamber resolution, the alignment and
the reconstruction precision.
\end{itemize}

As a result, the angular resolution is a complicated function, depending not
only on the muon momentum and on the amount of matter on the particle
trajectory, but also on the variables used in the event selection.
A good opportunity to characterise the angular precision and check the simulation
is given by the study of the space-angle distribution of two-track events
in the detector, called ``di-muon events'' in the following.
Muons coming from the decays of mesons originating from the early
stages of the shower development
are produced very high in the atmosphere and move along nearly parallel paths.
The angular separation of the two muons is therefore a good measure of the smearing introduced in their
direction by all the effects mentioned above. The results of the L3+C
simulation is compared with the obtained di-muon data.

\subsubsection{Di-muon analysis }
\label{sec-di-mu}

\begin{figure}[!ht]
\begin{center}
\includegraphics[bb= 1 1 550 360,clip,height=10 cm,width=17 cm]{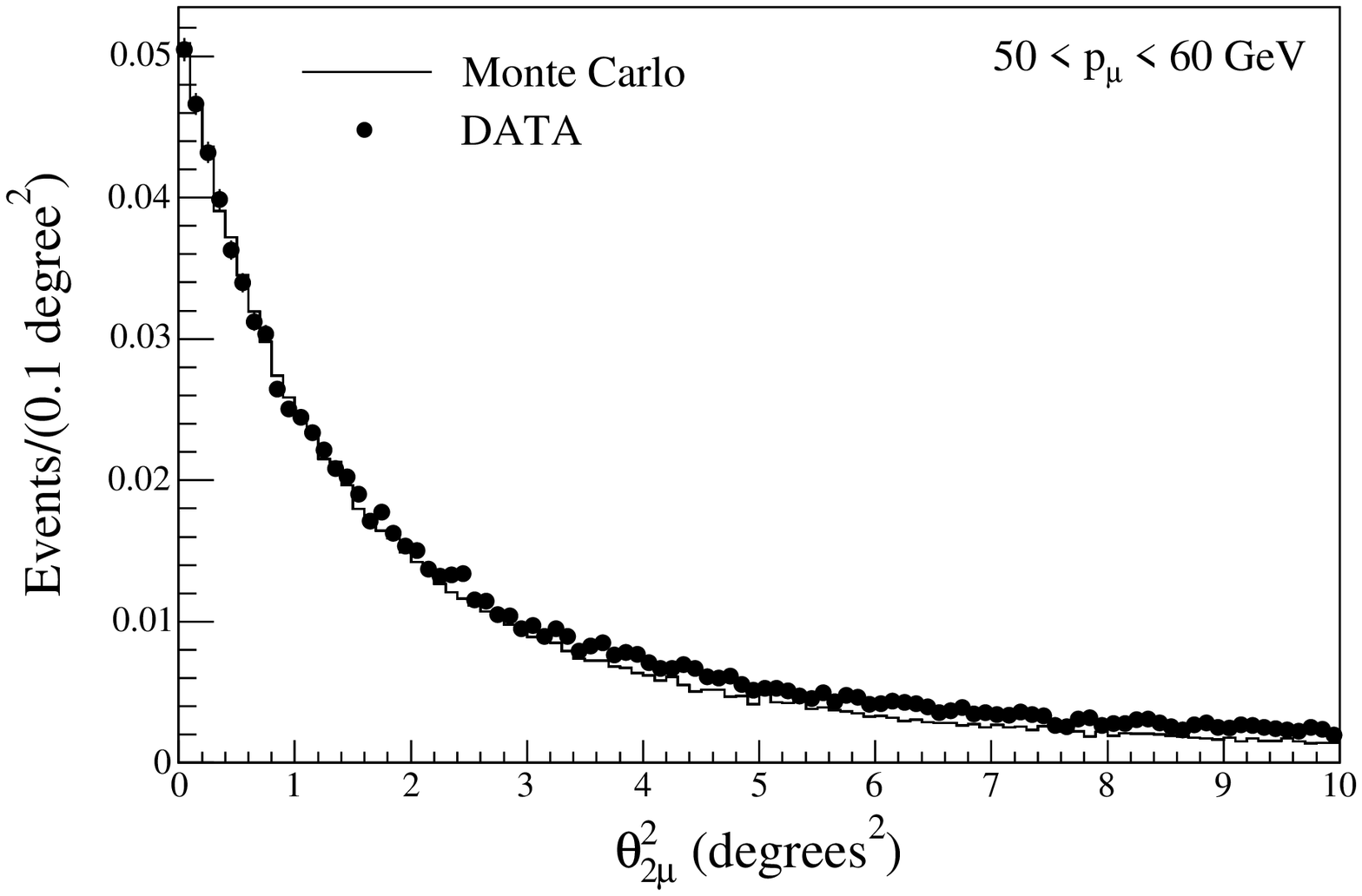}
\caption{\footnotesize{Distribution of the square of the angle between muons, $\theta^{2}_{2\mu}$, 
for data and Monte carlo. The distributions are normalized to unit area.}}
\label{di-mu_1}
\end{center}
\end{figure}

Di-muon events are selected with cuts requiring a minimum quality of the two tracks.
The main goal of the selection is to remove
fake di-muon events {\it i.e.} events with single muon split into two different tracks because
of reconstruction problems. For this purpose, a minimum separation between both
tracks is required. Events are further classified
into ``double-double'' ``double-single'' and ``single-single''
according to the number of subtracks for each
of the reconstructed track, a subtrack being defined for each octant crossed.
A muon momentum threshold, defined at the ground surface level, is also imposed on both
muons. Except for high muon momenta, large statistics
are available and results
are mainly dominated by systematic uncertainties. Events simulated
with the same sets of cuts are compared with experimental results.
Figure \ref{di-mu_1} is an example of such a comparison for all events
with muon momenta between 50 and $60\rm\;GeV$.

\begin{figure}[!ht]
\begin{center}
\includegraphics[bb= 1 1 560 380,clip,height=10 cm,width=17 cm]{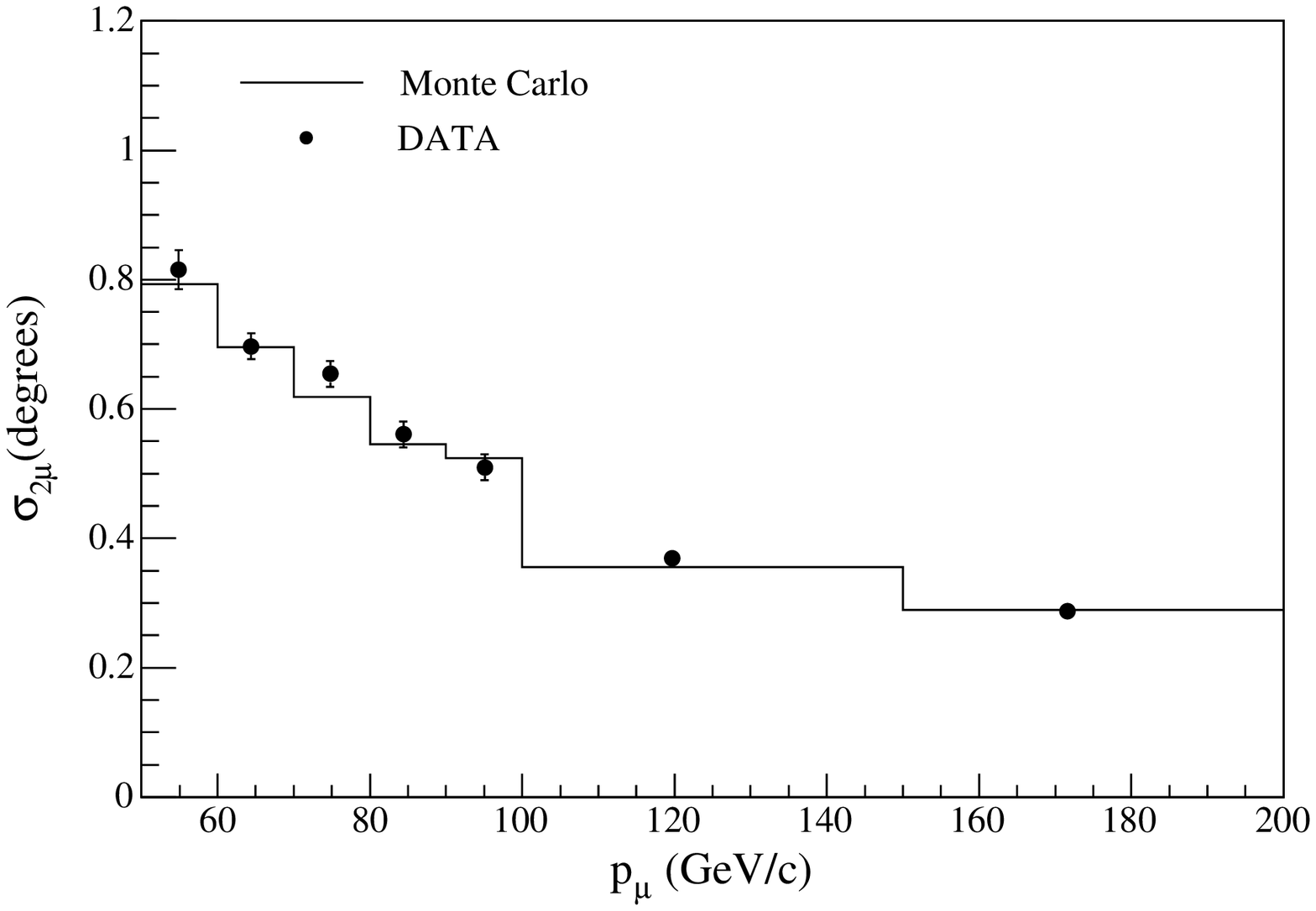}
\caption{\footnotesize{ Di-muon angular resolution versus the muon momentum.
}}
\label{di-mu_2}
\end{center}
\end{figure}

In the following, we define the di-muon angular resolution as
$\sigma_{2\mu} = {\rm HWHM}/1.17 $ with $\rm HWHM$ being the 
half-width at half-maximum of the distribution peak.
The observed and expected values of $\sigma_{2\mu}$ 
are shown in Figure \ref{di-mu_2} as a function of the muon energy 
for the whole di-muon sample.

Another check of the detector simulation follows from the study of
the angular resolution versus the amount of matter crossed by the muons
before reaching the detector.
The multiple scattering is the main factor in the
contributions to the angular resolution from different components.
A large range of matter thickness is available
by selecting events from the access-shaft direction (minimum energy loss)
or from large zenith angles (maximum energy loss).

\begin{figure}[!ht]
\begin{center}
\includegraphics[bb= 0 0 560 380,clip,height=10 cm,width=17 cm]{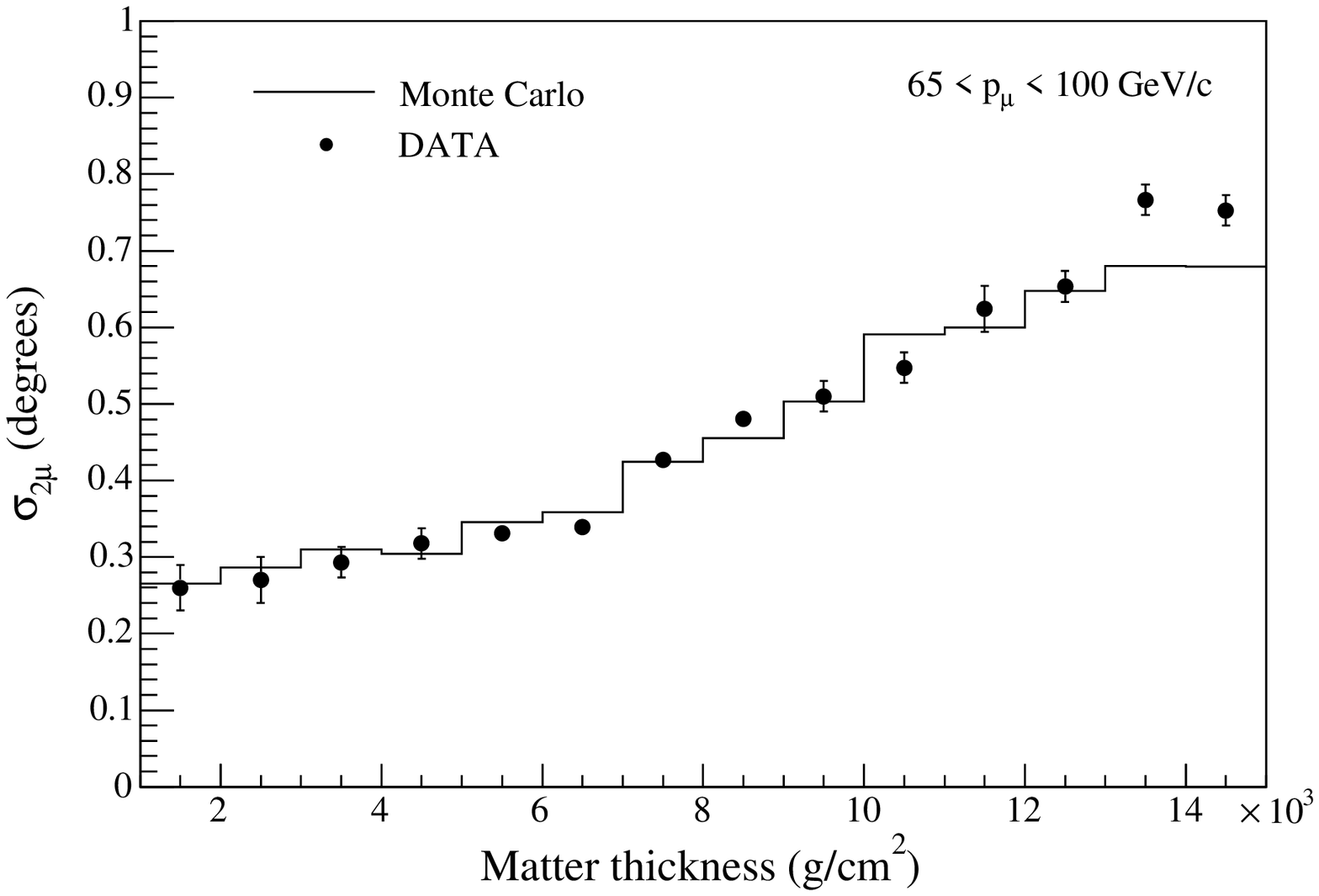}
\caption{\footnotesize{ Di-muon angular resolution versus 
matter thickness.
}}
\label{di-mu_3}
\end{center}
\end{figure}

Experimental results are compared with the detector simulation results
in Figure \ref{di-mu_3}. 
In all cases, except at large angles, the data is in a rather good agreement
with the results from the simulation.

\subsubsection{Angular resolution in the Moon shadow analysis}
\label{sec-sg-mu}

The present analysis uses single-muon events whose
angular resolution cannot be deduced from di-muon data.
In addition, as the distribution of both type of events inside the air shower
are quite different, they do not have the same acceptance. 
The good agreement observed between data and 
Monte Carlo in the di-muon analysis gives confidence in
the use of the simulation to extract the angular resolution in two ways:

\begin{itemize}
\item The angular resolution can be directly extracted from the event deficit in the Moon
direction. As already mentioned this is a unique opportunity for
an angular resolution measurement and the results will be compared with the simulation. 
\item Alternatively, the angular resolution can be fully
constrained with the help of the simulation of single muon events.
The point-spread function for each momentum bin is determined using the
same track selection and the same acceptance as for the selected Moon events.
The corresponding angular distributions can then be used in the shadow simulation as
smearing sources for the angular resolution.
Systematic uncertainties are at the level of 5\,\% or less, 
better than the angular resolution which can be extracted from the di-muon data.
These systematic uncertainties are
estimated from the comparison of data and simulation in the di-muon results and
from the studies of different production models in the simulation of air
showers with the CORSIKA program.
\end{itemize}

 \subsection{Earth magnetic-field and calculations of the deflection}

Two models are investigated to describe the geomagnetic field, a simple dipole
model and the International Geomagnetic Reference Field model (IGRF) \cite{IGRF}. In the latter, the
geomagnetic field is commonly expressed as the gradient of a scalar potential 
which can be expanded in terms of spherical harmonics. The IGRF consists of
a series of values of the coefficients in the expansion based on direct
measurements of the geomagnetic field.
In fact, the first terms of the expansion can 
be identified with the field produced by a dipole located at the centre of the
Earth. The contributions of the other terms can be considered as
perturbations of the main dipole field.
Quantitative differences between results of both models have been studied.
The main difference is a shift of the southern magnetic pole. As a
consequence, differences at the 10\,\% level
are observed at the L3+C location concerning the field intensity. 
Differences on the magnetic field direction
can reach $5^{\circ}$ in the part of the sky where the Moon is visible by
the detector. As a result, calculated deflections can differ at the same level.
The amount of deflection is overestimated by the dipole model in the largest
part of the sky and reaches 10\,\% for a $1\rm\;TeV$ proton.
Consequently only the IGRF model is used in the following.

Two coordinate systems are used. The first is based on the local horizon.
The zenith angle $\theta_{\rm Z}$ and the
azimuth angle $\alpha_{\rm Z}$ are determined from the precise position of the detector. 
The second is an equatorial system
using the Earth rotation axis as basis for the two coordinates: declination, $\delta$,
and right ascension, $RA$. 

The deflections of the particle trajectories in the magnetic field is described in an additional
coordinate system.
During its way from the Moon to the Earth, a particle of charge $Z$ and momentum $p$
is subject to the 
Lorenz force in the field $\overrightarrow{B}$ and the 
deflection  $\Delta\theta$ is linked to the
particle path $l$, as:

\begin{eqnarray}
\Delta\theta(\rm mrad) & = & 0.3 \cdot \frac{Z}{p({\rm TeV})} \cdot \left|\int
\overrightarrow{B}(\rm Tesla)\times d\overrightarrow{l}(\rm m)\right|.
\label{eq:defl}
\end{eqnarray}

\begin{figure}[!ht]
\begin{center}
\includegraphics[height=15 cm,width=15 cm]{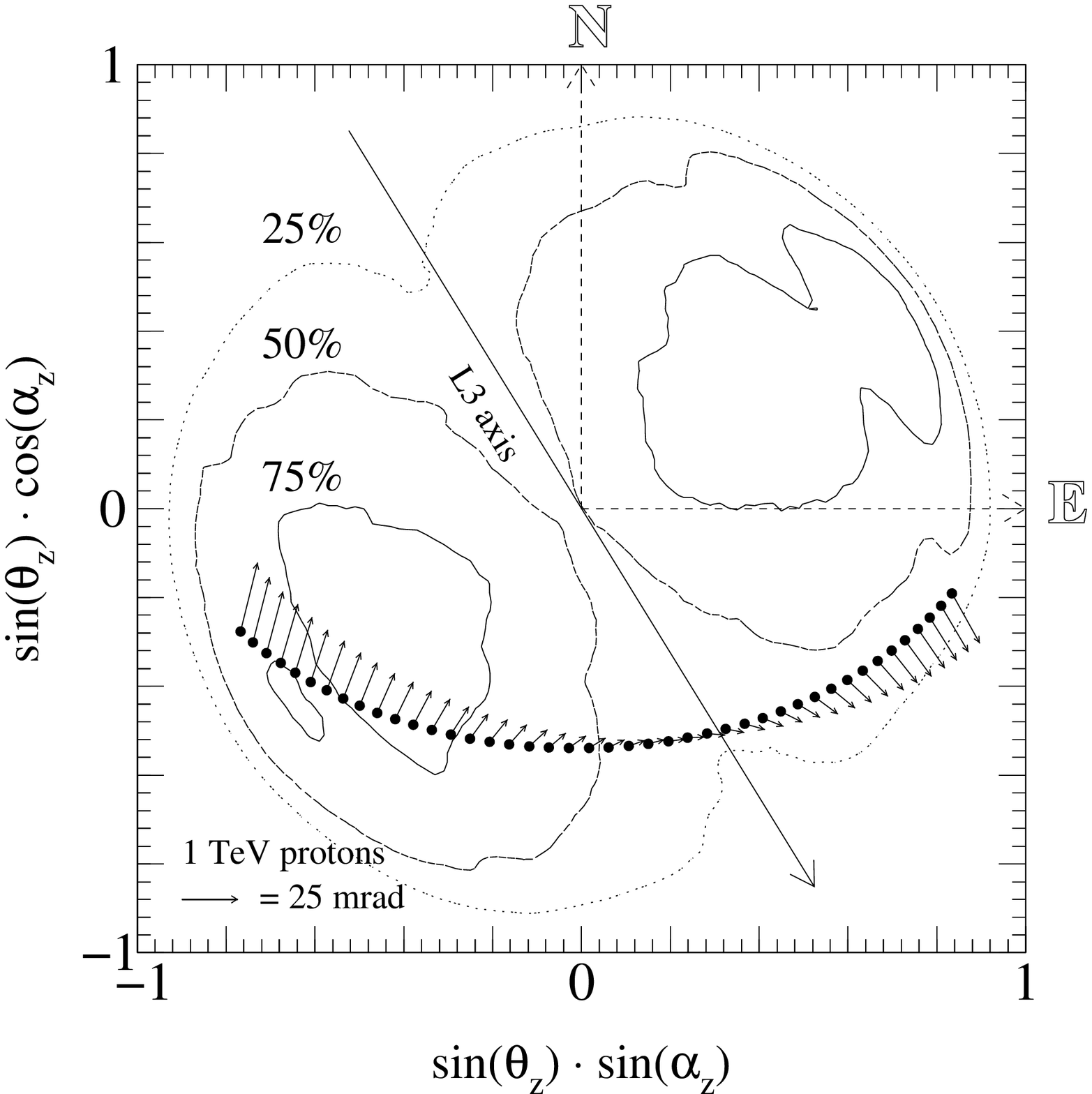}
\caption{\footnotesize{The Moon passing through the L3+C acceptance. The acceptance is determined by the
particular structure of the drift chamber and scintillator assembly.
The contour lines correspond to an observed cosmic-ray flux of
respectively $75\%$, $50\%$ and $25\%$ of the maximum of the flux.
A Moon transit is indicated
with dots. For each dot, the geomagnetic-deflection direction and amplitude
for a $1\rm\;TeV$ proton is indicated by an arrow.}
}
\label{def_2}
\end{center}
\end{figure}

Depending on their incident direction at the top of the atmosphere, charged
particles traverse different field regions. Thus, for a given
particle, the angular deflection is a function of the incident 
direction, the charge and the momentum of the particle. This can be used to
establish a deflection map that gives, for a given momentum, the amount
of deflection and its direction.
Figure \ref{def_2} shows a Moon transit above the sky as seen by the L3+C
detector. Each point corresponds to one direction in the sky as computed
from the zenith and azimuth angles.
During a Moon transit in the sky, the direction of deflection strongly 
depends on the Moon position but the dependence on the 
momentum is rather small. This leads to the definition of a coordinate
system defined for each Moon position in the sky, with coordinates
$\theta_{\rm H}$ and $\theta_{\rm V}$
respectively parallel and orthogonal to the direction 
computed for a particle with a given primary momentum (here a $1\rm\;TeV$ proton).
The indexes H and V stand here respectively for `horizontal' (parallel deflection) and `vertical'
(perpendicular to deflection).
In this way, magnetic deviations will shift the Moon shadow image
along the parallel direction and the shape in the other direction 
will mainly depend on the angular resolution.

 \subsection{Primary cosmic-ray composition}
\label{sec-primary}

Energy spectra for various elements,
up to a few hundreds TeV for protons and a few tens
TeV per nucleon (TeV$/$N) for heavy elements, have been measured
with the use of balloon and satellite experiments. The proton
spectra obtained by different experiments are in reasonable agreement.
Results are considerably scattered for other
elements, a consequence of limited statistics and uncertainties in
the energy calibration.

A compilation of available data \cite{wiebel} proposes the
following fit for the flux of particles:
\begin{eqnarray}
  \Phi = \Phi_{0} E^{-\gamma_{A}}
\end{eqnarray}
with $E$ the energy per nucleus in TeV.
The power index for $\rm He^4$ is smaller than the one for protons, and therefore
the contribution of $\alpha$ particles increases at high energy.
However, recent results from the RUNJOB \cite{runjob}, AMS \cite{AMS} and BESS \cite{BESSc} collaborations seem to
invalidate such behaviour, with a common power index $\sim~2.8$ for
proton and helium spectra.

The composition of primary cosmic rays plays an important role for the Moon shadow. 
It acts remarkably differently for experiments
using EAS arrays, Cherenkov or $\mu$ detectors. For the first
two methods, the measured signal is proportional to the total energy $E$ of the
primary. The third method, characterised by the $\mu$ momentum
threshold, is sensitive to the primary energy per nucleon $E_{\rm N}=E/A$.

Muons with energy $E_{\mu}$ are produced by nucleons of a minimum energy 
$E_{0}$ with 
$E_{0} \simeq E_{\mu}$ and thus by nuclei with energy $E > AE_{0}$. 
In a nucleus, all $A$ nucleons may contribute to the interaction.
Let us call $\sigma(E_{\mu},E_{\rm N})$ the 
cross section for the production of a muon with energy $E_{\mu}$ by a nucleon 
with energy $E_{\rm N}$.
If the spectrum has an index $\gamma_{A}$ and the corresponding  
flux is $\phi_{A}$ at $1\rm\;TeV$, then the number of muons with energy $E_{\mu}$ 
produced by these nuclei is:
\begin{eqnarray}
N_{\rm A}(E_{\mu}) & = & A\int_{AE_{0}}^{\infty}\sigma(E_{\mu},E_{\rm N})
\phi_{A}E^{-\gamma_{A}}\,dE.
\end{eqnarray}

If the energy spectrum of
all particles follows a power law with the same index $\gamma$ and 
the probability to yield a muon with energy $E_{\mu}$ does not depend on the
energy per nucleon above threshold,
the contribution of a nucleus with $A$ nucleons relative to the proton
contribution is :

\begin{eqnarray}
\frac{N_{A}(E_{\mu})}{N_{1}(E_{\mu})} & \propto & r_{A} A^{-\gamma+2},
\end{eqnarray}
where $r_{A}=\frac{\phi_{A}}{\phi_{1}}$ is the relative abundance of a nucleus
compared to the proton abundance at $1\rm\;TeV$. Using data
 from Reference \cite{wiebel}, it is found that muons originate at 75.8\,\% from protons, 
17.3\,\% from helium nuclei and 6.9\,\% from heavier nuclei.

According to equation (\ref{eq:defl}), for a mean primary energy $E_{\rm mean}$, the 
mean deflection angle is proportional to the ratio $\frac{Z}{E_{\rm mean}}$. For 
experiments sensitive to the total energy, like EAS arrays, the mean primary energy 
does not depend on $A$. Thus
\begin{eqnarray}
\langle \Delta\theta\rangle _{\rm EAS} & \propto & Z
\end{eqnarray}

The position of the Moon shadow depends only on $Z$, therefore EAS-array
experiments expect separate shadows for protons, helium and heavier nuclei.

For a muon experiment, the mean primary energy is proportional to $AE_{0}$. Thus
the mean deflection angle is:
\begin{eqnarray}
\langle \Delta\theta\rangle _{\mu} & \propto & \frac{Z}{A}
\end{eqnarray}

Muon experiments are sensitive to the ratio $\frac{Z}{A}$, which is equal to 1 
for protons, and from 0.5 to 0.4 for heavier nuclei. All shadows from helium and
heavier nuclei are almost at the same place.
Therefore, in the following, the primary flux for the
observed muons will be considered to be 75\,\% protons and 25\,\% helium nuclei.

 \subsection{Primary cosmic-ray energy spectrum}

\begin{figure}[!ht]
\begin{center}
\includegraphics[height=15 cm,width=15 cm]{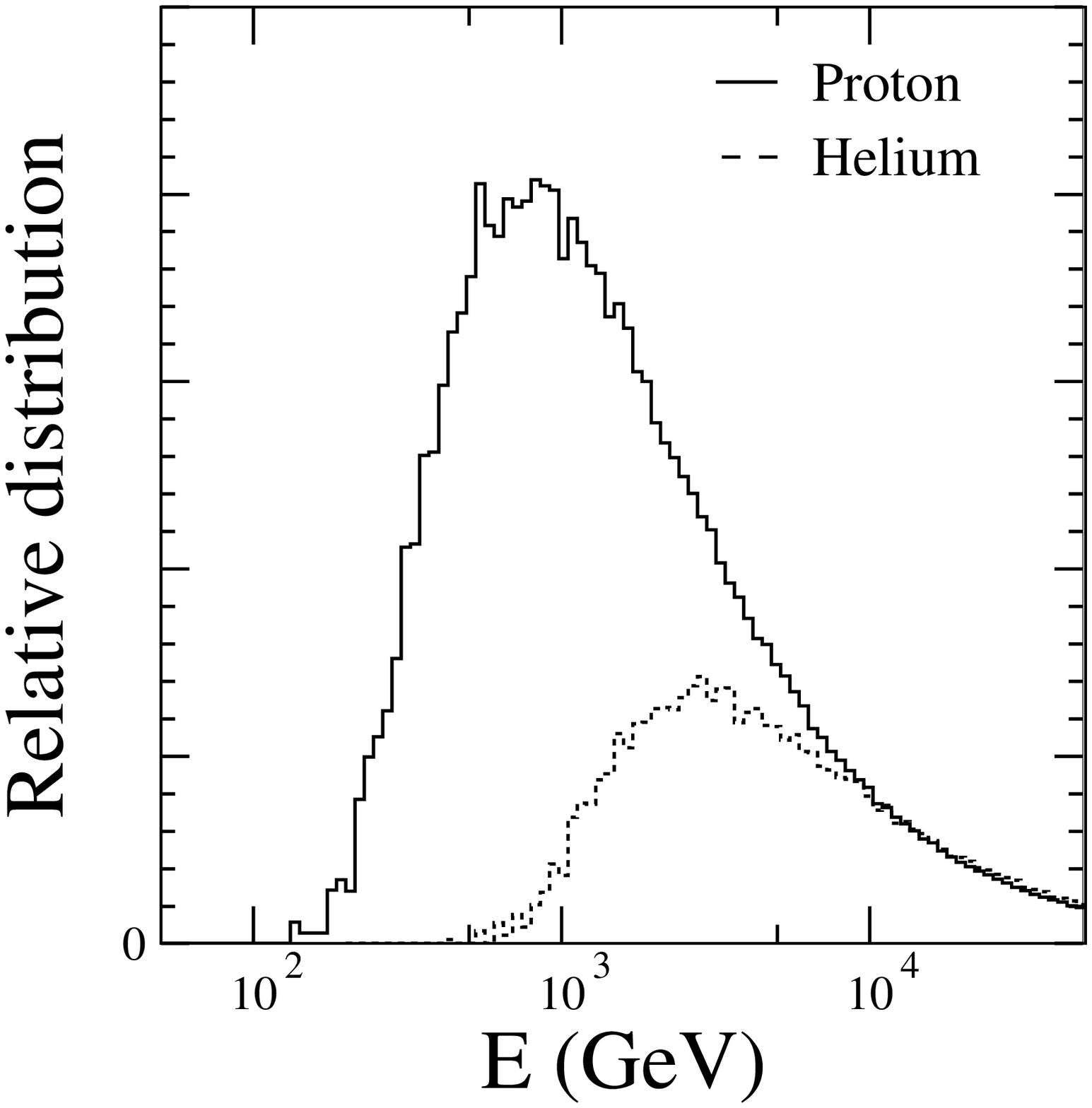}
\caption{\footnotesize{
Distributions of proton and helium primary energies for
$E_{\mu}= 100\rm\;GeV$. The normalisation between the distributions follows the chosen ratio:
$75\,\%$ of protons and $25\,\%$ of helium. The vertical scale is in arbitrary units.
}}
\label{primary_sp}
\end{center}
\end{figure}

There is an energy window for the observation of a magnetic-field
effect on the Moon shadow.
High primary energies ($\geq 10\rm\;TeV$) allow
ground level observations with large detectors
and relatively good statistics. However the magnetic deflection will be
small compared to the angular resolution, making the effect negligible
or, at most, appearing just as a small correction.
Low primary energies ($\sim 100\rm\;GeV$) are difficult to observe from the ground
and large deflections due to the
Earth magnetic field will dilute the shadow image and severely
limit the sensitivity.
L3+C has a good sensitivity
to muons from low energy primaries. Moreover, the muon energies are measured with 
good precision. 
The shadow effect can be observed using different ranges of muon energies,
thus selecting samples of different primary energy spectra.
For each observed muon energy $E_{\mu}$, a corresponding
primary energy $E$ with $E > E_{\mu}$
is obtained using the shower generation with
CORSIKA and the tracking of muons with the detector simulation.
Figure \ref{primary_sp} shows the expected proton 
and helium spectra associated with a detected muon with $E_{\mu}= 100\rm\;GeV$.
The maximum of the primary energy distribution is around $1\rm\;TeV$ for
protons and $4\rm\;TeV$ for helium nuclei.

\subsection{Moon-shadow simulation}
\label{shadow_simul}

\begin{figure}[!ht]
\begin{center}
\begin{tabular}{c c}
\includegraphics[height=8cm]{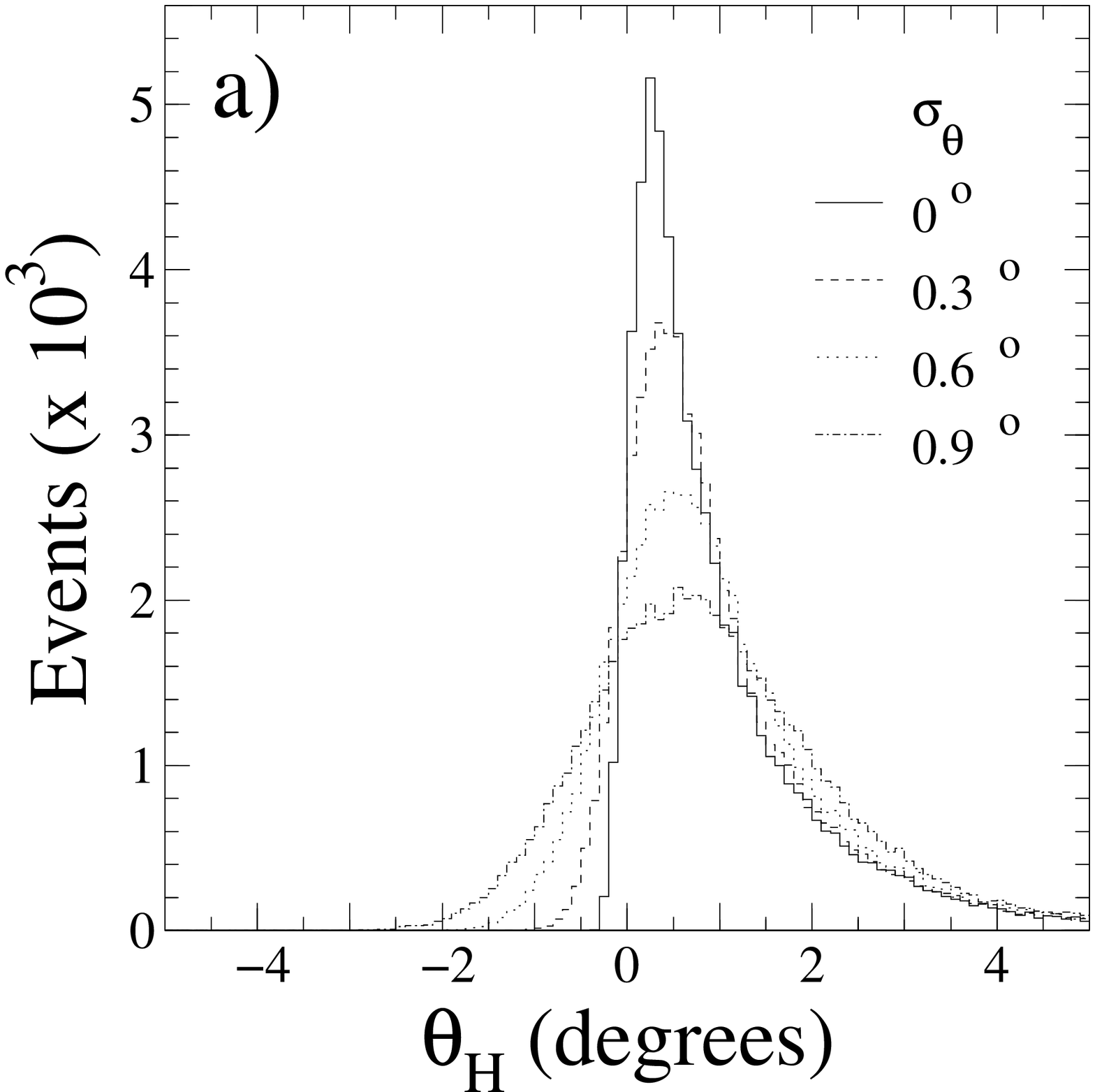} &
\includegraphics[height=8cm]{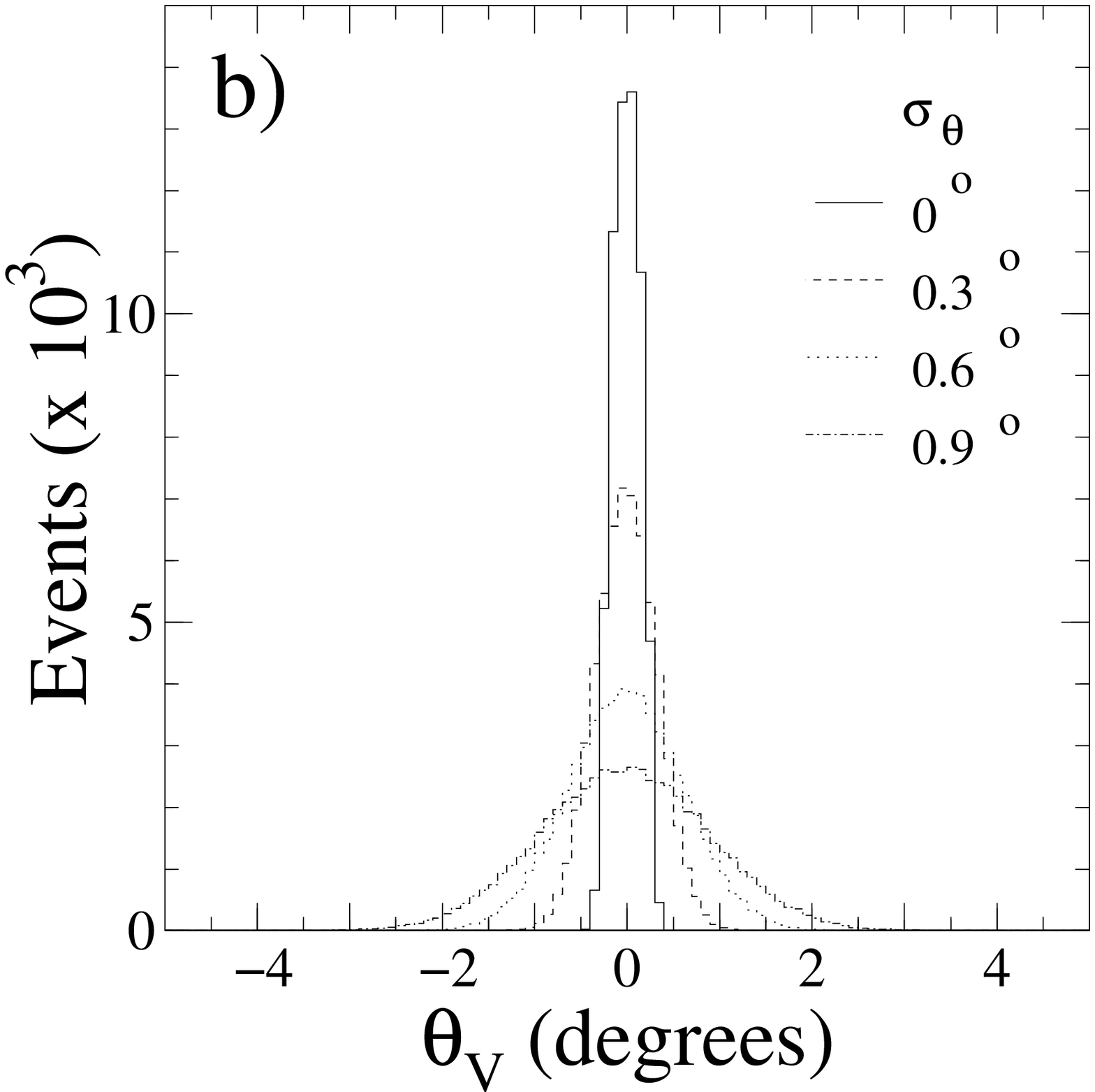} \\
\includegraphics[height=8cm]{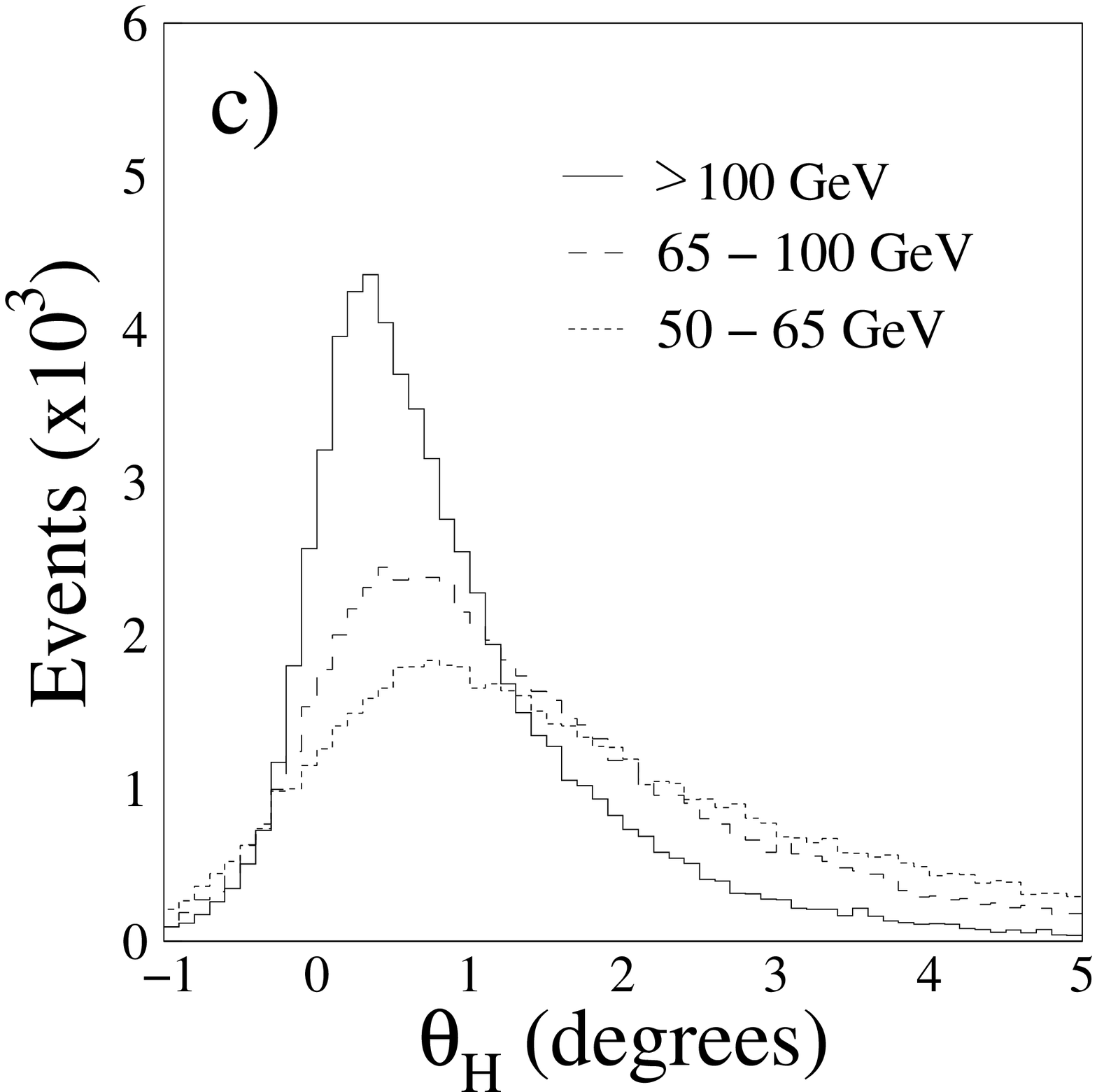} &
\includegraphics[height=8cm]{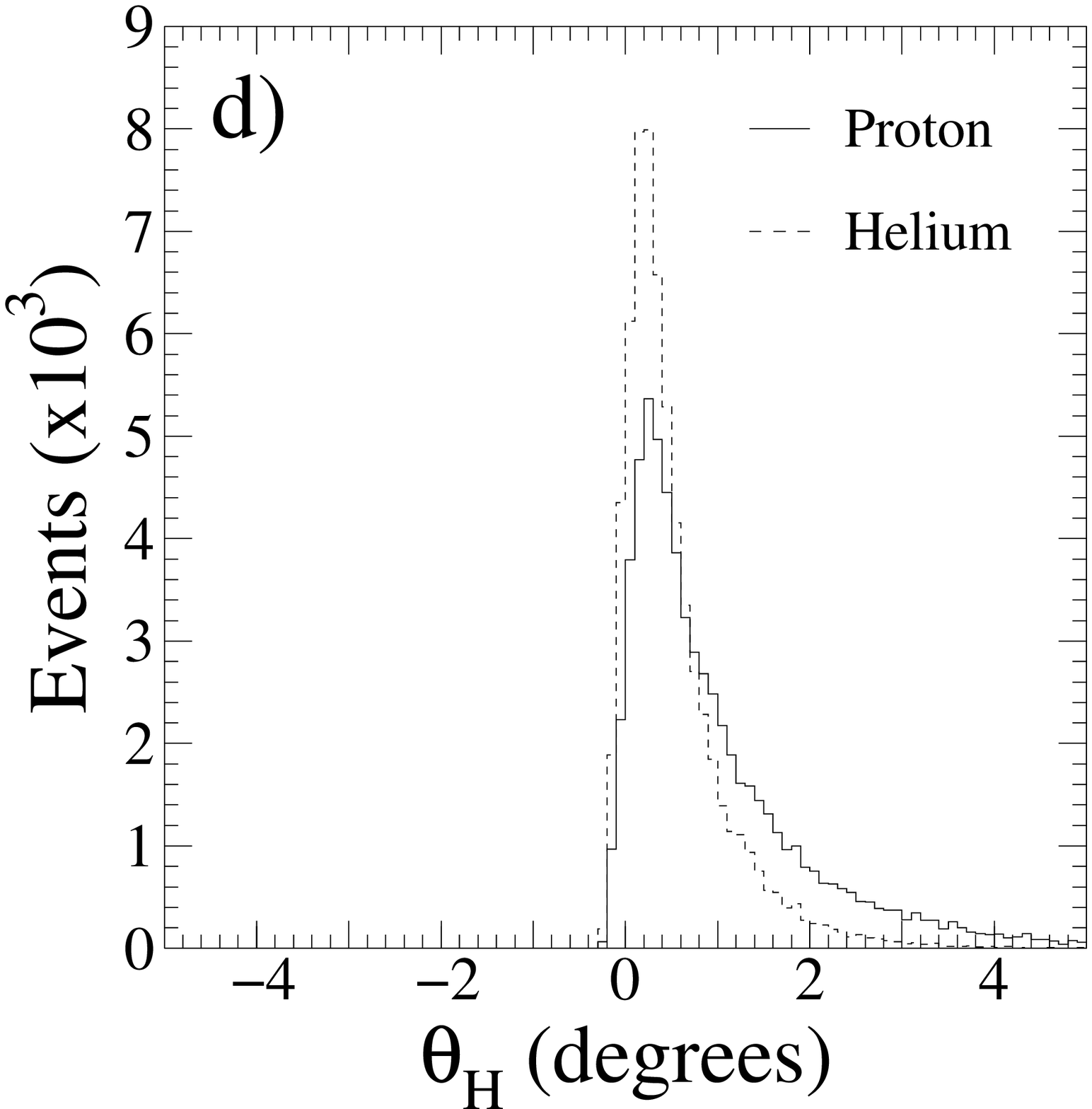} \\
\end{tabular}
\caption{ \footnotesize{
Examples of results of the simulation of the shape of the Moon shadow:
a) $\theta_{\rm H}$ for $E_{\mu}>100\rm\;GeV$ and for several 
hypothesis for the angular resolution,
b) $\theta_{\rm V}$ for $E_{\mu}>100\rm\;GeV$ and for several 
hypothesis for the angular resolution,  
c) $\theta_{\rm H}$ for three muon energy ranges,
d) $\theta_{\rm H}$ for proton and helium primaries for $E_{\mu}>100\rm\;GeV$.
All distributions are normalised to the same number of events.
}}
\label{smearmoon1}
\end{center}
\end{figure}

The simulation program tries to reproduce as closely as possible the conditions and
the parameters which significantly influence the observation of the
Moon shadow.
However to make the simulation more efficient, the particles are
assumed to be coming from the Moon surface: a positive signal is
simulated instead of a deficit. Also, for the same reason, particles are followed
backward instead of forward. They are originating from the detector 
and tracked through the Earth magnetic field up to the Moon.
The momentum and identity of the primary
particle are extracted from
the distributions described above.
For the angular smearing two methods are applied: either an arbitrary effective angular resolution is used,
or the momentum dependent angular information coming from the
experimental detector simulation.
Events are generated during a time span reproducing the experiment
running time and the detector acceptance. Some examples of
simulation results are shown in Figure \ref{smearmoon1}.

The shape of $\theta_{\rm H}$ becomes more symmetrical as
the angular resolution is worsening. In the orthogonal direction,
the smearing contribution is mainly coming from the angular resolution only. 
The expected signals for primary
protons and helium nuclei are 
largely overlapping and no extraction of
the helium contribution can be expected from the data.
Instead of the $\rm \Pap/\Pp$ ratio ($r_{\Pap/\Pp}$) mentioned before,
the analysis will try to get the ``$\Pap$ content'' as seen by L3+C,
 $r=\phi_{\Pap}/\phi_{\rm matter}$,
with $\phi_{\rm matter}$ the flux responsible of the observed deficit and assuming no anti-Helium contribution.
The $\rm \Pap/\Pp$ ratio itself can then be deduced from this result and the estimated 
proportion (75\,\%) of the deficit due to the proton flux relative to the the total matter flux,
as discussed in section \ref{sec-primary}.

\begin{figure}[!ht]
\begin{center}
\begin{tabular}{p{5.5 cm} p{5.5 cm} p{5.5 cm}}
\includegraphics[height=6 cm]{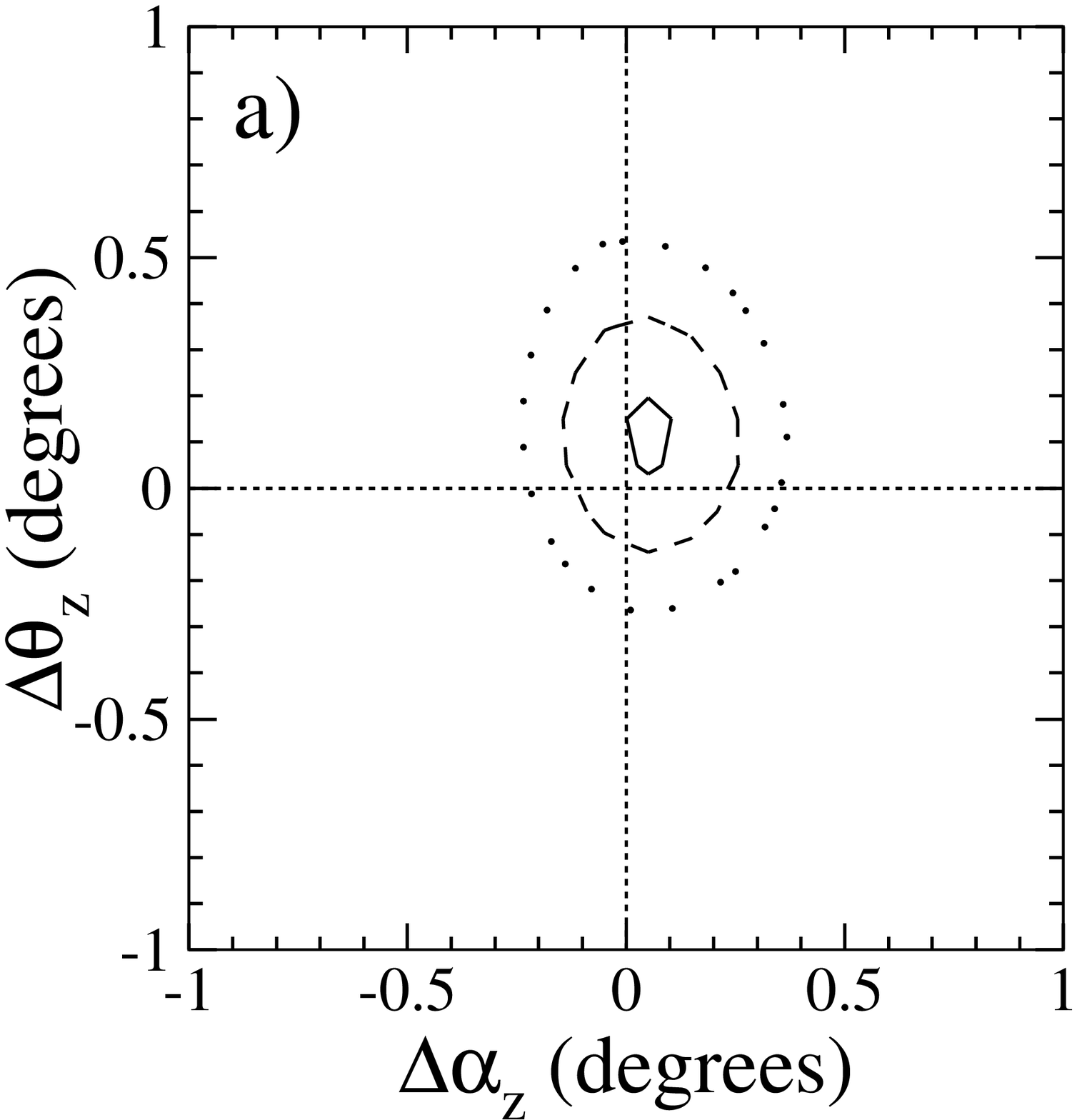} &
\includegraphics[height=6. cm]{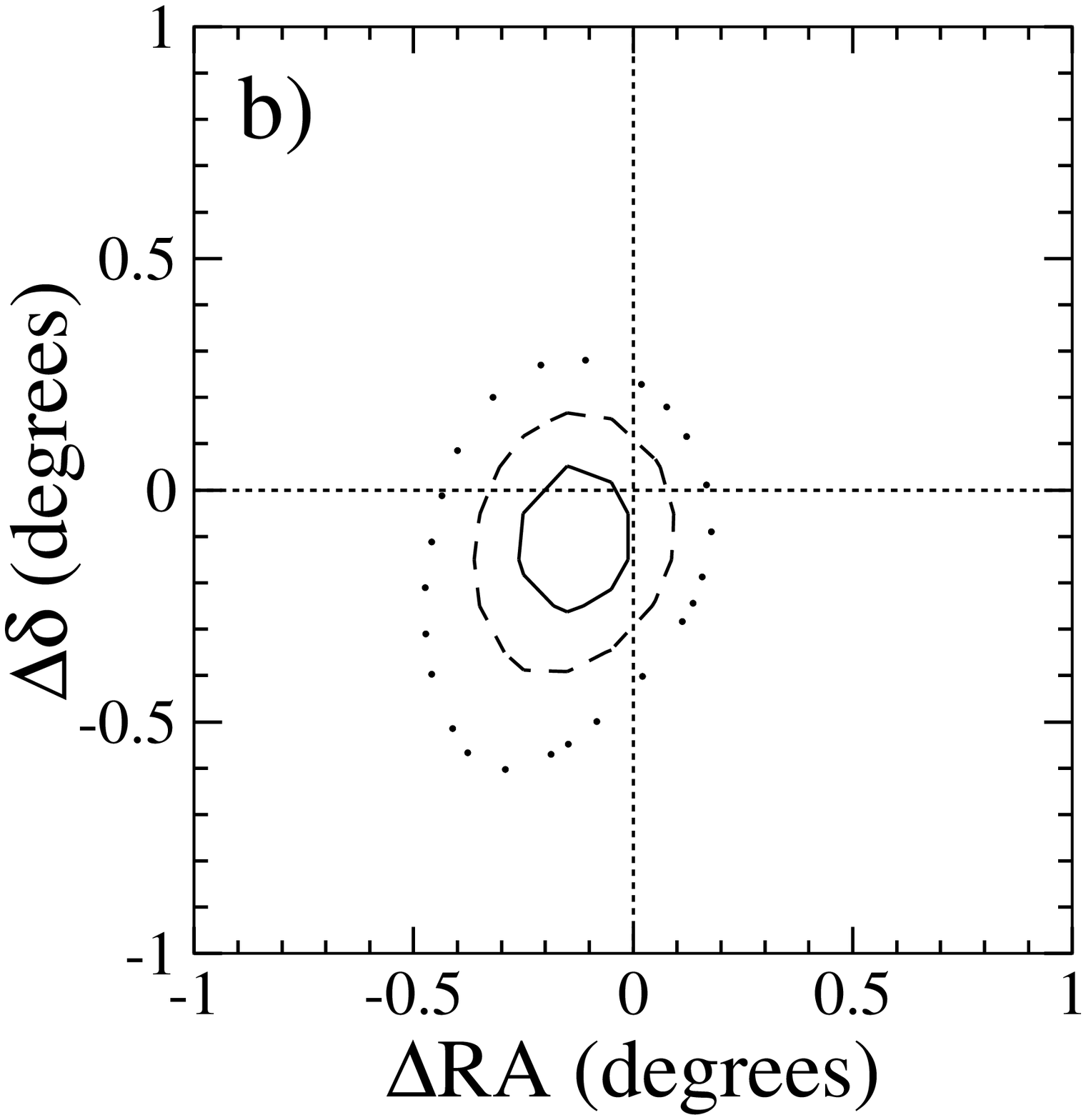} &
\includegraphics[height=6. cm]{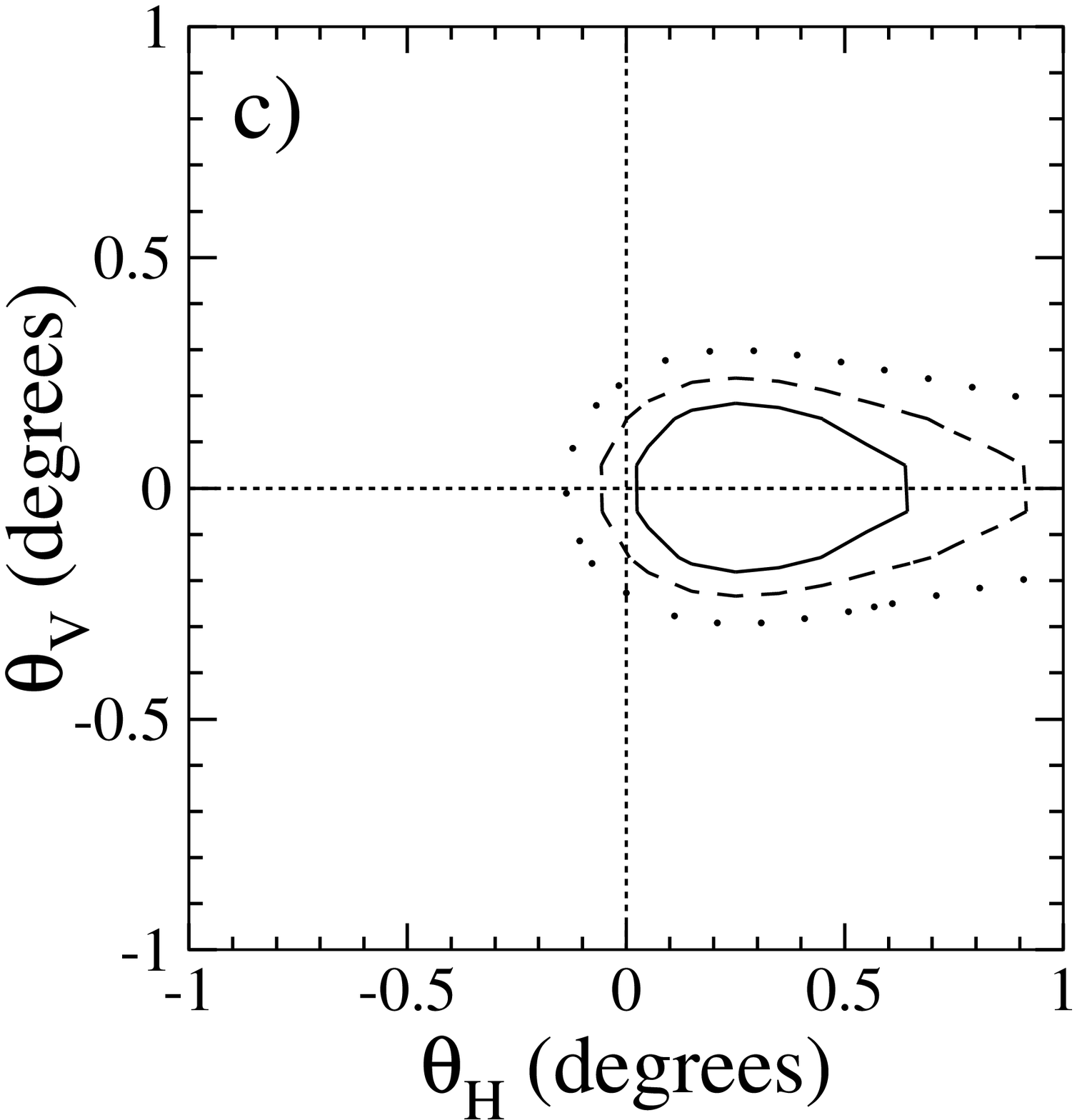} \\
\end{tabular}
\caption{\footnotesize{Contour map of the simulated Moon shadow 
for $E_{\mu}> 100\rm\;GeV$:
a) in the local coordinate system, b) in the
celestial coordinate system, c) in the deflection system. 
For each system a total of $6 \cdot 10^4$ events have been simulated.  
The solid contour line defines the domain containing more than 700 events per sky unit cell 
(of size $0.1^{\circ} \times 0.1^{\circ}$), and the dashed and dotted lines lower minimal numbers. 
The deflection system obviously records the largest amount of cells containing at least 700 events each.}}
\label{predic_coord}
\end{center}
\end{figure}

Figure \ref{predic_coord} shows the simulated Moon shadow as
it appears for $E_{\mu}>100\rm\;GeV$
in the three coordinate systems: local, equatorial and deflection.
Both the offset and the elongation due to the magnetic field are more 
visible in the deflection system. The search for a possible ``anti-shadow'' due
to antiprotons is therefore performed in this last system.

There are two ways to use the simulation results in this analysis,
either directly, or through a parametrisation.
In the first case, simulated distributions represent the expectation
values to be compared to the experimental data. In the second case, the
adjustment of the parametric function to the data  allows the extraction 
of the parameters. This is used to give a result on the
observed effective angular resolution.
The simulated shadow is parametrised as the product of
two functions, one for each direction:
\begin{equation}
f(x,y)\,=\,f_{\rm H}(x) \times f_{\rm V}(y),
\label{eq:shadow}
\end{equation}
$f_{\rm V}$ can be described as the projection on the vertical axis of the
two dimensional convolution of a disk (the Moon) with a
Gaussian distribution, corresponding to an effective angular 
resolution for the muon momentum range considered: 
\begin{eqnarray}
 f_{\rm V}(y) & = & \int_{-R_{\rm Moon}}^{+R_{\rm Moon}}\,\frac{2}{\pi
R_{\rm Moon}^{2}}\sqrt{R_{\rm Moon}^{2}-u^{2}}\,\,\,
\,\frac{1}{\sigma\sqrt{2\pi}}e^{-\frac{(y-u)^{2}}{2\sigma^{2}}}\,du.
\end{eqnarray}
For $f_{\rm H}$, no analytical description exists. However, it is found
that a sum of two Landau distributions approximates reasonably well
the shape in the direction parallel to the deflection:
\begin{equation}
f_{\rm H}(x)\,=\,(1-c)\cdot L_{1}(x,a_{1},b_{1})\,+\,c\cdot L_{2}(x,a_{2},b_{2}), 
\end{equation}
with the parameters $a$ and $b$ acting respectively on the width and the position
of the maximum of the distribution.
For each selected value of the angular resolution,
a number of events corresponding to 100 times the number of expected
deficit events in the experiment is simulated.
A good agreement between the simulated shape of the deficit
and the parametrisation is observed.

\section{The Moon data-samples }

To define the muon direction in the sky, both local and terrestrial
based coordinate systems are used.
The Moon position is computed using the
``SLALIB'' library subroutines \cite{slalib}.  The error on the
position calculation is estimated to be smaller than $0.01^{\circ}$,
much smaller than the angular resolution or the Moon radius. 
The Moon is the nearest and the only astronomical
object for which the position is significantly dependent on the
observation location on Earth. This parallax effect is taken
into account. It amounts to a few tenths of a degree.

Another effect is the change of the apparent
size of the Moon as seen from the Earth due to the variation of the 
centre-to-centre distance from the Earth to the Moon between perigee
and apogee.
Calculations show that the Moon angular radius, as seen from the Earth,
fluctuates between $0.25^{\circ}$ and $0.28^{\circ}$ 
during the whole data taking period.

The Moon declination is continuously changing inside a
range of around $\pm 20^{\circ}$.
In the local sky, the Moon follows a trajectory reaching a minimum zenith
angle of $25^{\circ}$ for the experimental running period. 

\begin{figure}[!ht]
\begin{center}
\includegraphics[height=18 cm,width=18 cm]{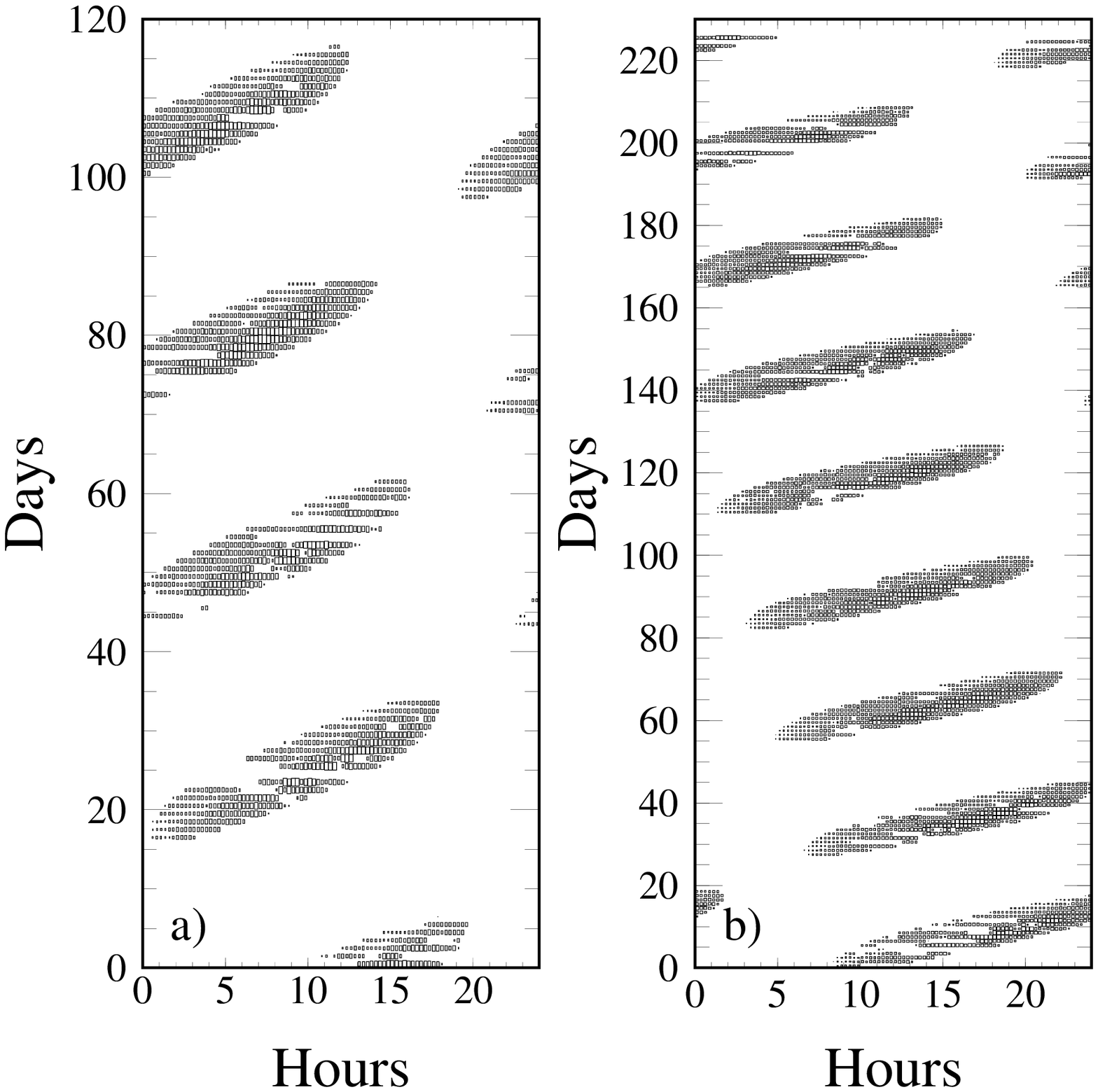}
\caption{\footnotesize{Moon acceptance for a) year 1999 and b) year 2000.
The plot shows the event arrival-time in days and hours. The running period
includes 5 lunar cycles in 1999 and 9 in 2000. White bands during cycles
correspond to periods in which either the detector or the data acquisition were not operational.
 }}
\label{accep}
\end{center}
\end{figure}

As the detector cannot be triggered above a certain zenith angle value,
the Moon is only available for certain periods
of time, each called a ``cycle'' in the following. Figure \ref{accep} shows
the Moon acceptance for the selected events for the two running years.
Data were accumulated for
five cycles in 1999 (73 transits) and nine cycles in 2000 (142 transits).
The Moon was available for a total
of 1557.5 hours. The corresponding data-acquisition live-time is 1188.7 
hours (76.3\,\%). 

\paragraph{Data selection and monitoring}

A first selection isolates events coming from the direction of the Moon.
For each event, the Moon position is computed in local coordinates
and the space angle $\theta$ with the muon track direction is deduced.
The track-reconstruction program requests at least one ``triplet''
(hits from 3 chambers in one
of the octants) and one scintillator hit. 
Two hits in one octant (called a ``doublet'') are also allowed if a momentum mesurement is possible,
but two doublets are rejected.   
This constraint leads to the reconstruction of only one third of the total
number of events collected. The following 
cuts are applied at this level:
\begin{itemize}
 \item Only one muon is reconstructed in each event.
 \item The angle with the Moon direction is less than $5^{\circ}$. 
 \item The muon momentum is more than $50\rm\;GeV$.
This cut is motivated by the fact that low energy muons
have little correlation with the direction of the primary cosmic ray.
 \item The Moon zenith-angle is less than $60^{\circ}$. Above this value 
    reconstruction becomes more difficult and the
    trigger efficiency is low.
 \item The backtracking of the measured muon track in the detector up to
   the surface is successful.
 \item Events with timing uncertainties, amounting to 0.2\,\% of the total, 
    are rejected as correct time information is needed
    to compute precisely the Moon position. 
\end{itemize}
When these cuts are applied, a total of $6.71 \times 10^5$ events are selected,
out of which $2.11 \times 10^5$ in 1999 and $4.60 \times 10^5$ in 2000.

To monitor the data,
some variables are carefully tested as a function of time. Among them
are the number of selected events, the number of high energy events,
the $\mu^{+} / \mu^{-}$ charge ratio, the proportion of muons including 2 subtracks
and the proportion of high quality subtracks in events.
Di-muon events are also used. They are taken
from directions along the Moon trajectory using the same runs
as in the Moon data. A total of more
than 50000 events are collected in this way.
The event selection requires that at least one of the muons has 2 subtracks
and that the minimum muon momentum is $50\rm\;GeV$.
The di-muon space angle distribution is monitored.

Stability with time is an indication
that no major problem perturbed the collection of data during all the running of the
experiment. No major instability is observed for the whole running period.

\paragraph{Background determination}

Searching for a point source (or shadow) entails the counting of the number
of events in an angular bin containing the possible signal
(signal bin) and comparing it to the number of background events expected in this bin. 
The background is a function $B(\theta_{\rm z}, \alpha_{\rm z},t)$ of the zenith
angle $\theta_{\rm z}$, the azimuth angle $\alpha_{\rm z}$, and the time t.

Apart from the $\theta_{\rm z}$ dependence due to the changing thickness of the atmosphere, 
there is a strong spatial dependence of the reconstruction
efficiency due to the arrangement of the muon chambers in octants and to the 
constraints imposed on the track reconstruction. 
The  background is therefore evaluated by counting events in regions 
that were exposed for the same amounts
of time to the same directions of the sky as the signal bin. 

The global rate is changing with time, due to modifications in the detector hardware
or in the detector environment (noise dependence, local atmospheric
temperature and pressure dependence). A correction
has therefore to be applied. In general time and spatial
angular dependence are independent and the spatial 
acceptance is nearly constant. If not, one has to consider sufficiently
small time slices so that this is valid.

\begin{figure}[!ht]
\begin{center}
\includegraphics[height=17 cm,width=17cm]{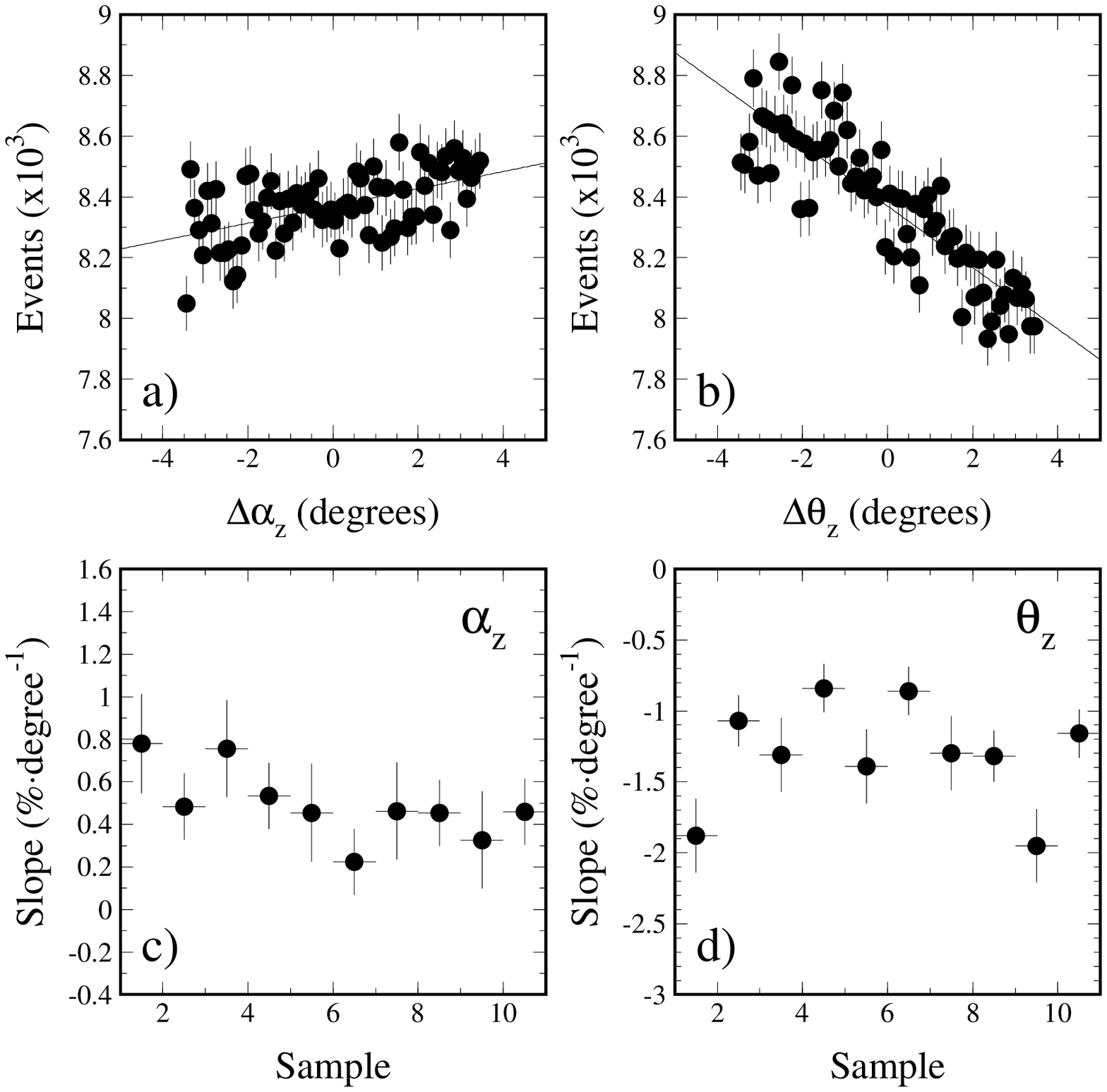}
\caption{\footnotesize{
   Number of events versus a) the azimuth and b) the zenith direction, using
     the merged data from the 10 samples. The origin is taken at the real or
     ``fake Moon'' position.
  c) Slope parameters extracted for
  the 10 considered samples in the azimuth direction and 
  d) in the zenith direction. The first 2 bins are from
     signal and the remaining from background.
}}
\label{az_z_var}
\end{center}
\end{figure}

The background is determined by measuring
the number of events due to ``fake Moons''. These are bins which  
cross a given region in the sky either earlier or later than the
signal bin in different runs. Averaging background samples on both sides
of the signal bin removes effects of changes in the event rate which
are linear in time. Four ``fake Moons'' are used, one hour
and two hours before and after the real Moon position.
When considering both running years, ten samples are available for the
background evaluation (the two signal samples and eight background samples).

The projections on the azimuth and on the zenith axes are shown in
Figure \ref{az_z_var}a and Figure \ref{az_z_var}b.
In these examples, the merged data from all samples are used.
The distributions are fitted by a straight line. 
The fitted slope parameters from the individual samples 
are shown in Figures
\ref{az_z_var}c and \ref{az_z_var}d for the azimuth and zenith directions respectively.
There is only a slight positive variation of the rate as function of the azimuth angle
and all the samples give statistically
compatible results. The variation with the zenith angle is more important.
The rate is decreasing for large zenith angles, a consequence of the acceptance.
Moreover sample-to-sample fluctuations are larger.

Each sample is fitted by a plane in the azimuth-zenith
coordinate system. The expected background
density at the origin (real or ``fake Moon'' position) is computed,
expressed by the number of events in the Moon solid angle, $\Omega_{\rm Moon}$, 
for the given live-time of the whole experiment.
Fluctuations between samples are at the level
of a few percent, much greater than the statistical uncertainties. 
Therefore systematic uncertainties due to
live-time errors and acceptance or rate changes with time dominate.
Following the hypothesis mentioned above that time and spatial
angular dependence are independent, a further normalisation correction-factor
based on the overall number of events for each sample can be applied.
This is obtained from the total number of events inside an
annulus around the nominal Moon position $(3^{\circ} < \theta < 5^{\circ})$.
After this normalisation, differences are at the 0.1\,\% level.
When applying the above procedure, there is no significant difference in the
evaluation of the background using different samples.
The systematic uncertainties in the knowledge 
of the background rate at the Moon position are negligible compared to
the statistical uncertainties on the signal.
The result averaged for the two years is
$\rm 542 \pm 0.6\;events/\Omega_{\rm Moon} \simeq 2366 \pm 3\;events/deg^{2}$
for $E_{\mu} > 100\rm\;GeV$
and $\rm 677 \pm 1.0\;events/\Omega_{\rm Moon} \simeq 2956 \pm 4\;events/deg^{2}$
for $65\;{\rm GeV}< E_{\mu} < 100\;{\rm GeV}$.
\section{Event-deficit analysis}

\begin{figure}[!ht]
\begin{center}
\includegraphics[height=15 cm,width=15 cm]{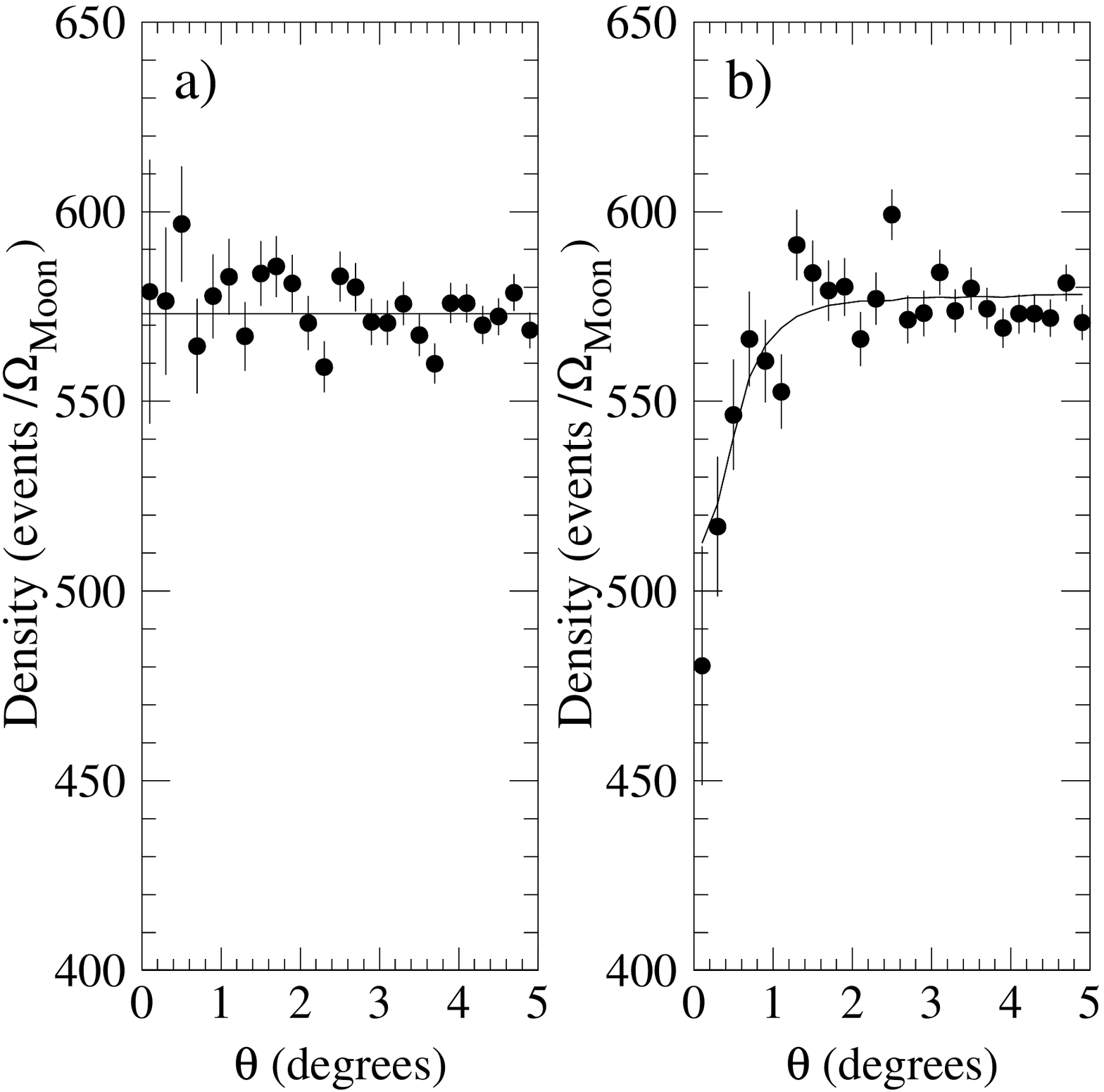}
\caption{\footnotesize{Angular distributions  
for events$/$Moon-solid-angle with $E_{\mu} > 100\rm\;GeV$,
(a) for a ``fake Moon'' shifted 1 hour behind its real position,
(b) using the correct Moon position.
Solid lines are the results of the simulation, including the angular resolution
deduced from the study of di-muon events.
}}
\label{1d4}
\end{center}
\end{figure}

In the local coordinate system, evidence for a cosmic ray deficit introduced by the Moon
is observed using a single angular variable.
The number of events shows a
linear increase with increasing solid angle. Therefore the density
of events is generally considered. 
In absence of any signal, plots must show
a flat distribution. The plots of Figure \ref{1d4} show the results for
 $E_{\mu} > 100 {\rm\;GeV}$
with a ``fake Moon'' shifted 1 hour behind its real position along
its trajectory (Figure \ref{1d4}a) and with the Moon at its nominal
position (Figure \ref{1d4}b). In this last plot, a clear deficit of events in the
first few bins is observed. This is attributed to the shadowing effect of the Moon.

Uncertainties in pointing and the influence of the geomagnetic
field are also contributing to the shape of the deficit. 
The extraction of the experimental angular resolution
and the measurement of the $\rm \Pap$ content in primary cosmic rays are
not possible on this one dimensional distribution.
In the following, a maximum likelihood method is used to disentangle
the various factors.
An interesting property of the 
``deflection coordinate system'', which is based on the deflection induced by the
Earth magnetic field, is to concentrate the Moon shadow deficit along one
axis, thus optimising the signal density. Therefore the analysis will be
performed in this coordinate system with the two projection angles
$\theta_{\rm H}$ and $\theta_{\rm V}$. 
Studies to investigate the effect of the muon momentum threshold on the deficit
lead to the definition of two samples,
a ``high energy (HE)'' sample for
$E_{\mu} > 100 {\rm\;GeV}$ and a ``low energy (LE)'' sample for $65 {\rm\;GeV} < E_{\mu} < 100 {\rm\;GeV}$.

\begin{figure}[!ht]
\begin{tabular}{c c}
\includegraphics[height=8 cm]{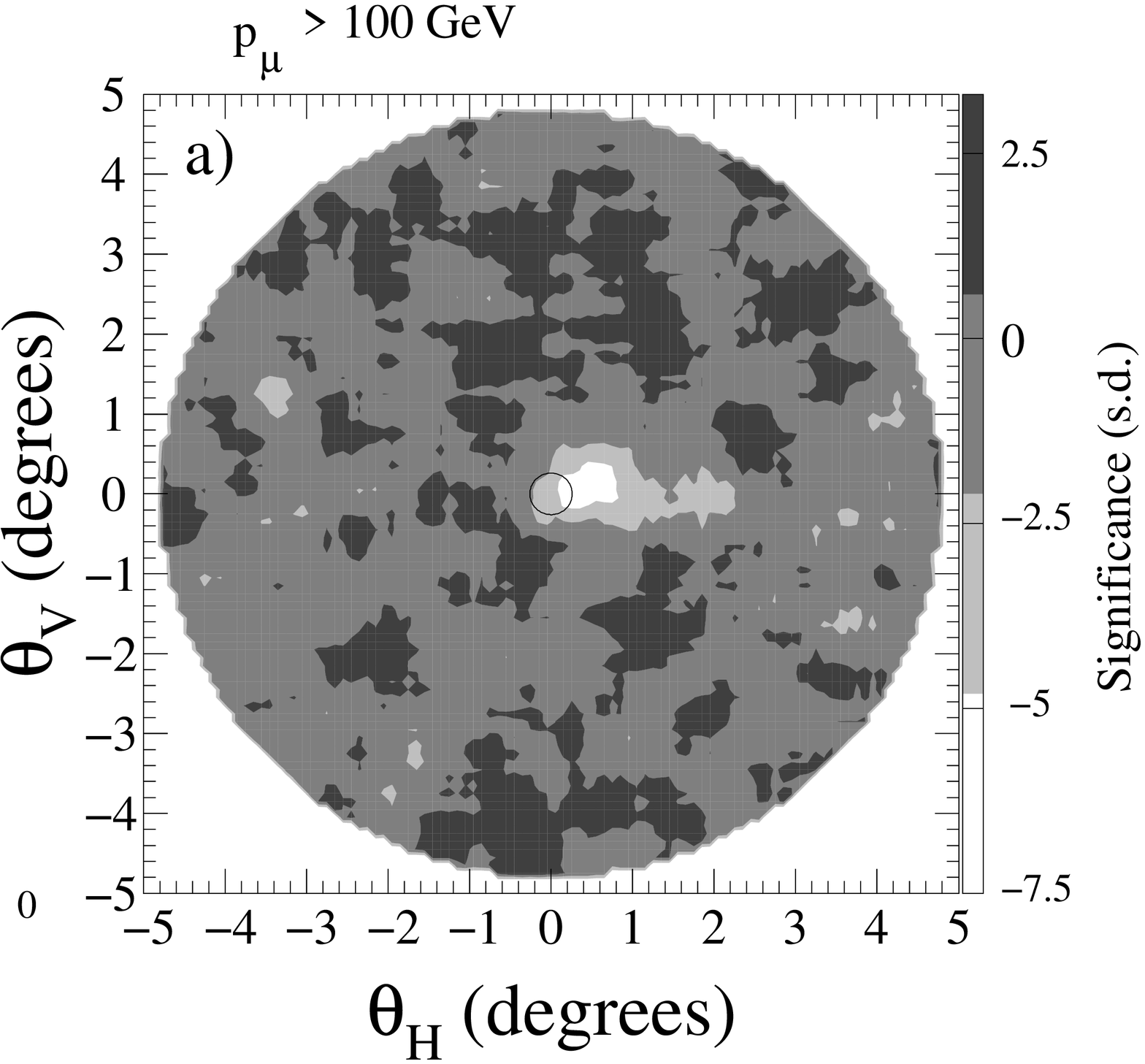} &
\includegraphics[height=8 cm]{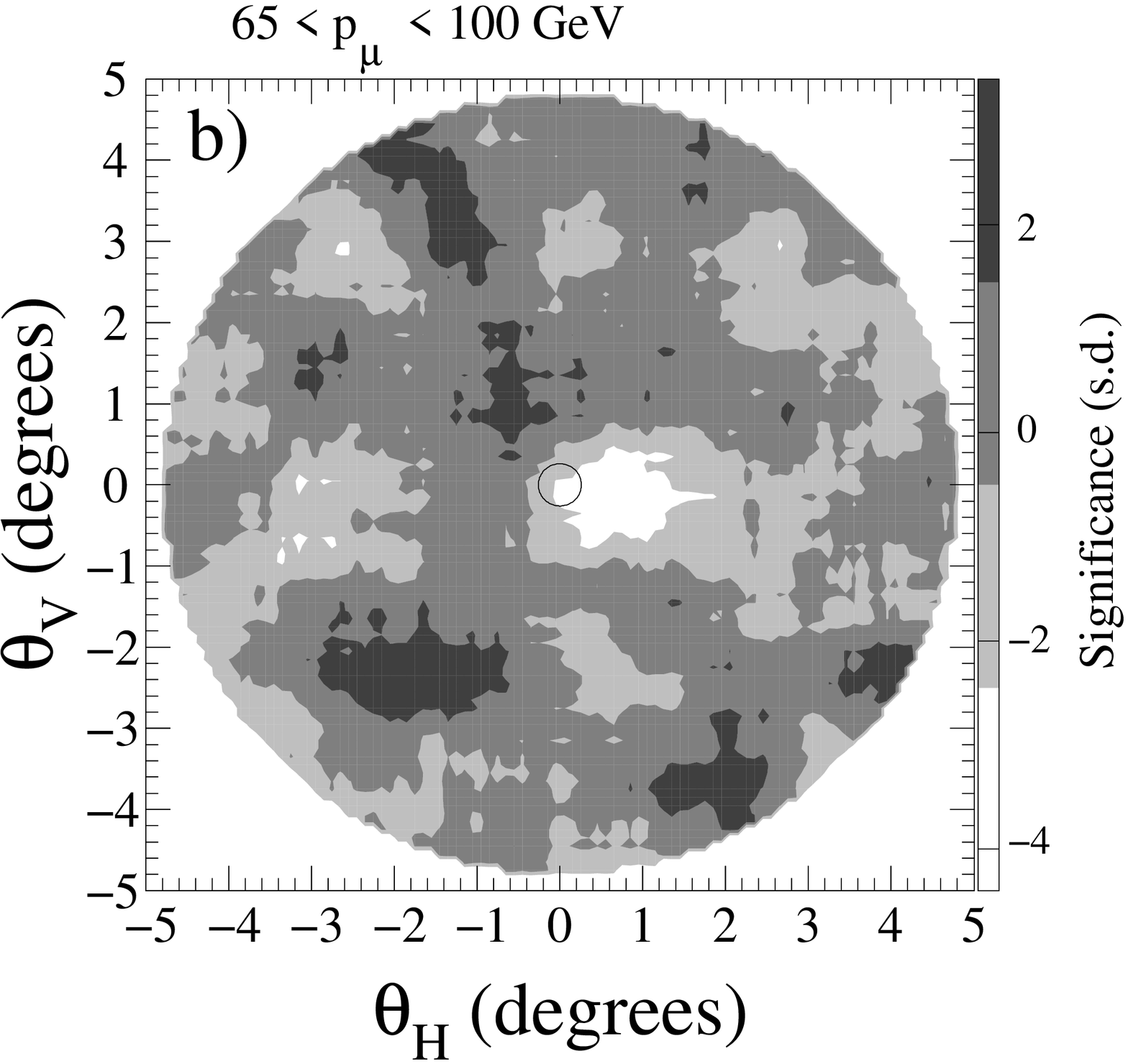} \\
\end{tabular}
\caption{\footnotesize{Results obtained in the deflection system for:
  a) the High Energy sample, b) the Low Energy sample.
In both cases, smoothing techniques have been applied. 
A circle indicates the true position of the Moon.
The vertical grey scale shows the significance in standard deviation units; negative values 
correspond to an event deficit.}}
\label{data_h}
\end{figure}

Figure \ref{data_h} shows ``shadow''plots concerning the data for both samples.
Smoothing techniques are used.
The way the background is computed is described later.
For the analysis, ``raw'' spectra are used.
A binning of $0.1^{\circ}$ is chosen in each direction.

The shape of the shadow is more elongated in the case
of the LE sample and its position is shifted further.
The position and the shape of the Moon-related deficit 
mainly depend on the magnetic deflection undergone by the primary
particle associated with each muon
of the sample and on the effective angular resolution $\sigma$.
Both effects are taken into account in the simulation.
The effective angular resolution includes the muon production-angle inside the air shower,
the multiple scattering in the molasse above the detector and the intrinsic
angular-resolution due to muon-chamber resolution, alignment and reconstruction.
The smearing due to the multiple scattering is the main component of 
$\sigma$.
A parametrisation of the simulation output for each value of 
$\sigma$ and for each sample was described in \ref{shadow_simul}.
It is used in the maximum likelihood fit to allow the extraction of $\sigma$
as a free parameter.

\paragraph{Analysis procedure}
\label{anal-proc}

The probability to have $n_{i,j}$ events in bin $(i,j)$ of the $\theta_{V} \, - \, \theta_{H}$ distribution, when $g(x_{i},y_{j})$
events are expected is discribed by Poisson statistics. The logarithm of the likelihood function can be written as:
\begin{eqnarray}
\ln\,\mathcal{L}\,\equiv\,\sum_{i,j=1}^{N}\,\ln(P_{i,j})\,=\,\sum_{i,j=1}^{N}\,
[\,{n_{i,j}}\ln(g(x_{i},y_{j}))-g(x_{i},y_{j})-\ln(n_{i,j}!)\,].
\end{eqnarray}
For practical reasons, the likelihood function is normalised and a
minimum $\Delta \ln\mathcal{L}_{\rm m}$ is searched as a function of the parameter values.
The two-dimensional distributions are the result of the combination of 3 different components:
\begin{itemize}
\item{a smooth background, which can be fitted with a plane,}
\item{the proton and helium deficit introduced by the Moon shadow,}
\item{the antiproton deficit if any.  }
\end{itemize}

The simulation provides a description of the proton deficit.
A similar description is used for the antiproton deficit.
However, in the deflection coordinate system, the shadow position is inverted with respect to the Moon centre.
Also the shape of the antiproton shadow will differ from that of the protons, due to a possible
different power index $\gamma$ of the energy spectrum.
Thus the most general description of the data is:
\begin{eqnarray}
g(x,y) & = & \underbrace{u_{x}x+u_{y}y+u_{z}}_{\rm background}\,-\,\frac{N_{\rm miss}}{1+r}
[0.75\underbrace{f_{1}(x-x_{0},y-y_{0},\sigma)}_{\rm \Pp\ {\rm deficit}}+
0.25\underbrace{f_{2}(x-x_{0},y-y_{0},\sigma)}_{\rm He\ deficit}+{}\nonumber\\
 &  & {}+~r\underbrace{f_{3}(x_{0,\rm\overline{p}}-x,y_{0,\rm\overline{p}}-y,\sigma_{\rm\overline{p}})}_{\rm\Pap\ {\rm
 deficit}}],
\label{eq:mlh}
\end{eqnarray}
where $f_{1}$, $f_{2}$ and  $f_{3}$ are the shadow functions defined by equation (\ref{eq:shadow}),
respectively for protons, helium nuclei and antiprotons.
The parameters that can vary for the maximum likelihood fit
are the angular resolutions $\sigma$, the 
positions of the deficits $(x_{0},y_{0})$, the number of missing events $N_{\rm miss}$
and $r$, the $\rm \Pap$ content.  
The parameters $u_{x}, \, u_{y}, \, u_{z}$ describe the background.

The influence of different power indexes $\gamma$ ($-$1.7 to $-$3.7) of the antiproton energy spectrum  
on the final limit of the antiproton to proton ratio presented in this paper is studied, 
as well as a re-analysis of the data assuming also different values of $\gamma$.
Variations of less than 20\,\% with respect to a simplified
model, where the power index is assumed identical for protons and antiprotons are obtained.
The interpretation of this conclusion may be explained by the fact that for a steep
antiproton spectrum the deflection of most antiprotons is relatively large, but the muon statistics 
small, due to the lower average primary energy.
In the case of a flatter spectrum the deviation is small, but the muon statistics larger.
The two effects cancel each other, providing a negligible influence on the limit of the 
$\rm \overline{p}/p$ ratio.
For simplicity we present only the result of the analysis with the assumption of equal
power indexes, reducing thus the number of free fit parameters to 8 ($f_{3} \, = \, f_{1}, \,
x_{0, \rm \overline{p}} \, = \, - \, x_{0}, \, y_{0, \rm \overline{p}}, \, = \, - \, y_{0}, 
\sigma_{\rm \overline{p}} \, = \, \sigma $). 

Instead of trying to extract directly the 8 parameters with the simplfied equation (\ref{eq:mlh}), the
analysis proceeds in several steps.

\paragraph{Background estimation}

A first determination of the background parameters is performed using the
ring-data defined as $ 3^{\circ} < \theta < 5^{\circ}$, where $\theta$ is the angle
between the muon and the nominal position of the Moon.
The cut excludes the cells in the proton and antiproton deficit regions.
The event density at the nominal Moon position is known at the
0.3\,\% level. This uncertainty corresponds to the statistics used for its
determination. 

No significant changes in the parameter values are found when equation (\ref{eq:mlh})
is applied to the whole angular range and all the parameters are considered free.

\begin{table}[!ht]
\begin{center}
\caption{\footnotesize{
Results obtained in the fit of the matter deficit.
}}
\vspace{4.mm}
\small
\begin{tabular}{|c|c|c|c|c|} \hline 
               &                            &                 &                         &               \\
               &  HE                        &  HE             & LE                      & LE            \\
parameter      & measured                   & expected        & measured                & expected      \\
               &                            &                 &                         &               \\
\hline
               &                            &                 &                         &                \\
{$\it x_{0}$}  & 0.33 $\pm$ 0.08            & 0.26            & 0.53 $\pm$ 0.13         & 0.48           \\
       	       &                            &                 &                         &                \\
{$\it y_{0}$}  & 0.05 $\pm$ 0.05            & 0.0             & $-$ 0.10 $\pm$ 0.08     & 0.0            \\               
       	       &                            &                 &                         &                \\
FWHM           & $1.07^{\,+0.07}_{\,-0.04}$ & 1.03            & 1.80 $\pm$ 0.15         & 1.87           \\             
       	       &                            &                 &                         &                \\
$N_{\rm miss}$ & $575^{\,+97}_{\,-87}$      & 546 $\pm$ 5     & $536^{\,+133}_{\,-127}$ & 683 $\pm$ 6    \\
       	       &                            &                 &                         &                \\
Significance   & 8.3 {$\it s.d.$}           & 8.0 {$\it s.d.$}& 5.5 {$\it s.d.$}        & 5.8 {$\it s.d.$} \\ 	      
       	       &                            &                 &                         &                \\
\hline 
\end{tabular}
\label{result_7par}
\end{center}
\end{table}

\paragraph{The Moon shadow analysis }

\begin{figure}[!ht]
\begin{center}
\includegraphics[height=8cm,width=8cm]{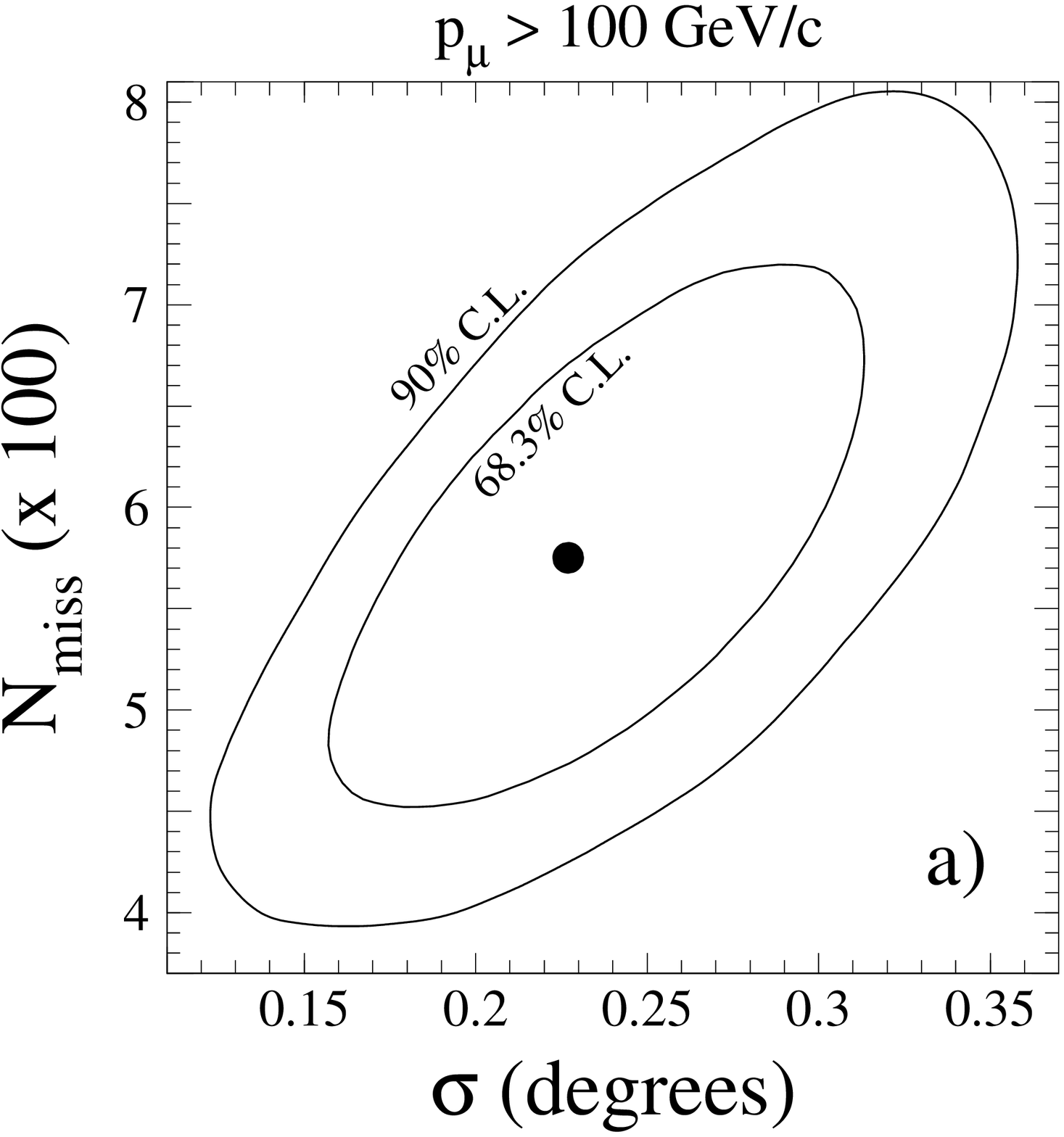}
\includegraphics[height=8cm,width=8cm]{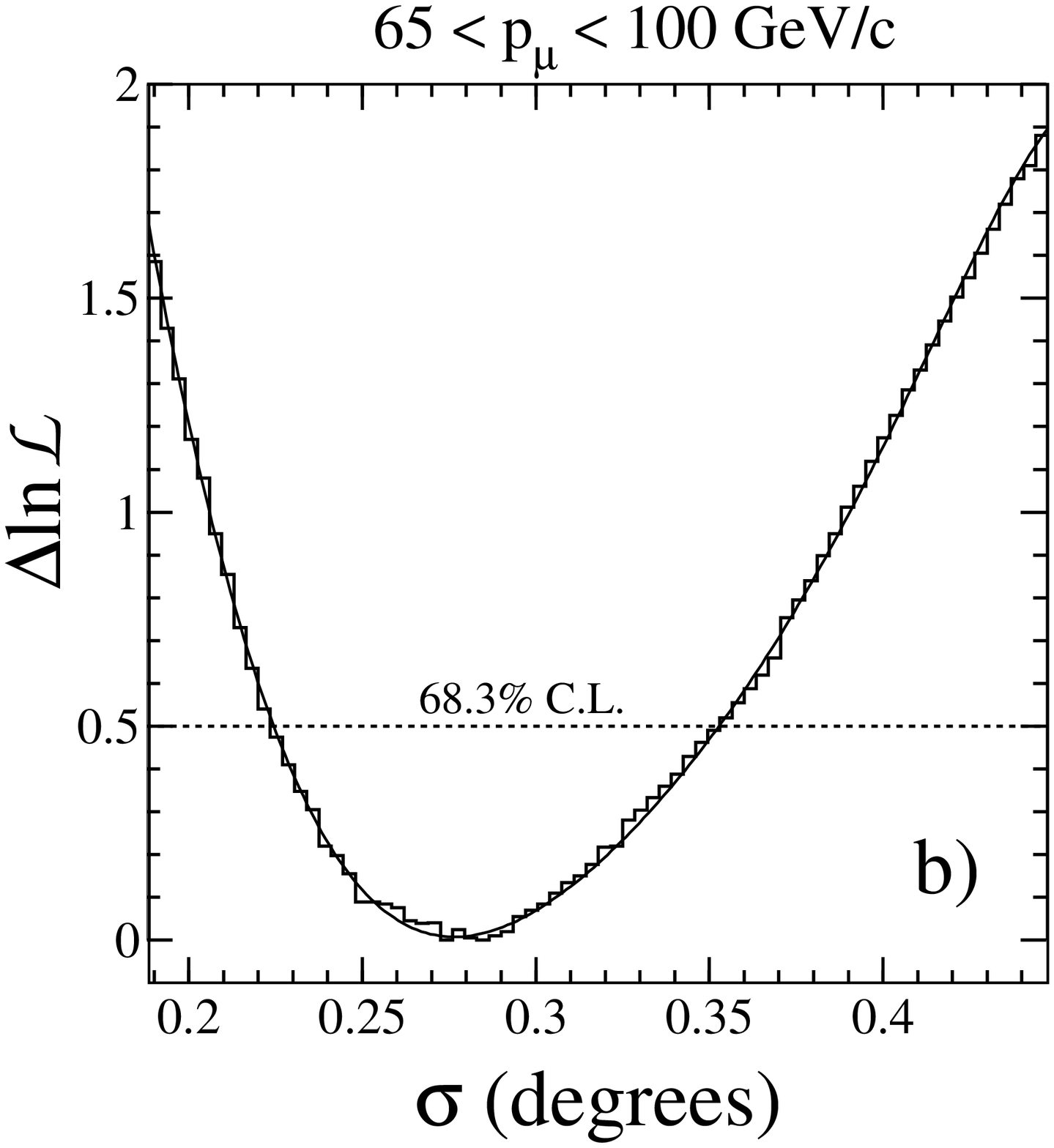} \\
\caption{ \footnotesize{
a) Two-dimensional 68\,\% and 90\,\% C.L. contour curves for the parameters $N_{\rm miss}$ and
$\sigma$ for the HE sample,
b) $\Delta \ln\mathcal{L}$ versus $\sigma$ for
the LE sample. $N_{\rm miss}$ has been constrained.
The dashed line for $\Delta \ln\mathcal{L} = 0.5$
is used to determine the $68.3\%$ central confidence interval.
}}
\label{likevssigma}
\end{center}
\end{figure}

In this step no antiprotons are supposed in the primary flux.
The effective angular resolution of the detector,
together with the pointing uncertainties,
are obtained from the observation of the matter deficit only.
Results of the  maximum likelihood  fit for the two samples are given in Table \ref{result_7par}.
As an example, two-dimensional 68\,\% and 90\,\% confidence level contour curves
for the parameters $N_{\rm miss}$ and
$\sigma$ are shown in Figure \ref{likevssigma}a for the case of the HE sample.

Pointing errors are
given by horizontal or vertical offsets between data and simulation
in the determination
of the deficit position $x_{0}$ and $y_{0}$.
Both values are small ($\leq 0.1^{\circ}$).

Values related to the absolute position, $x_{0}$, and to the
extension of the deficit (FWHM, full width at half maximum) 
in the horizontal direction $\theta_{\rm H}$
show clearly a momentum dependence.
In the $\theta_{\rm V}$ direction,
no shift is observed and the width is mainly the result of the effect of
$\sigma$.

\begin{figure}[!ht]
\begin{center}
\includegraphics[height=8.5cm,width=8.5cm]{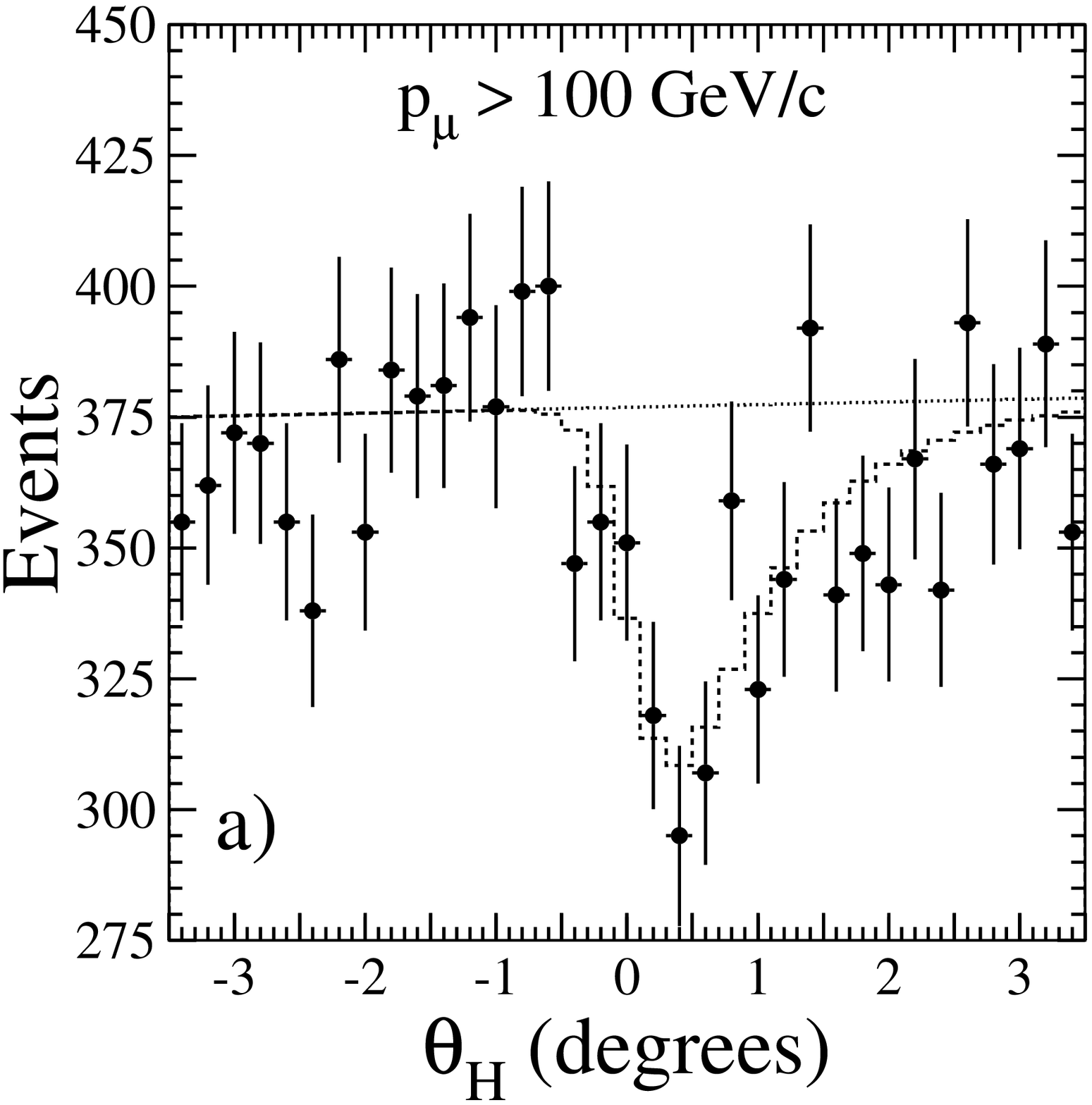} 
\includegraphics[height=8.5cm,width=8.5cm]{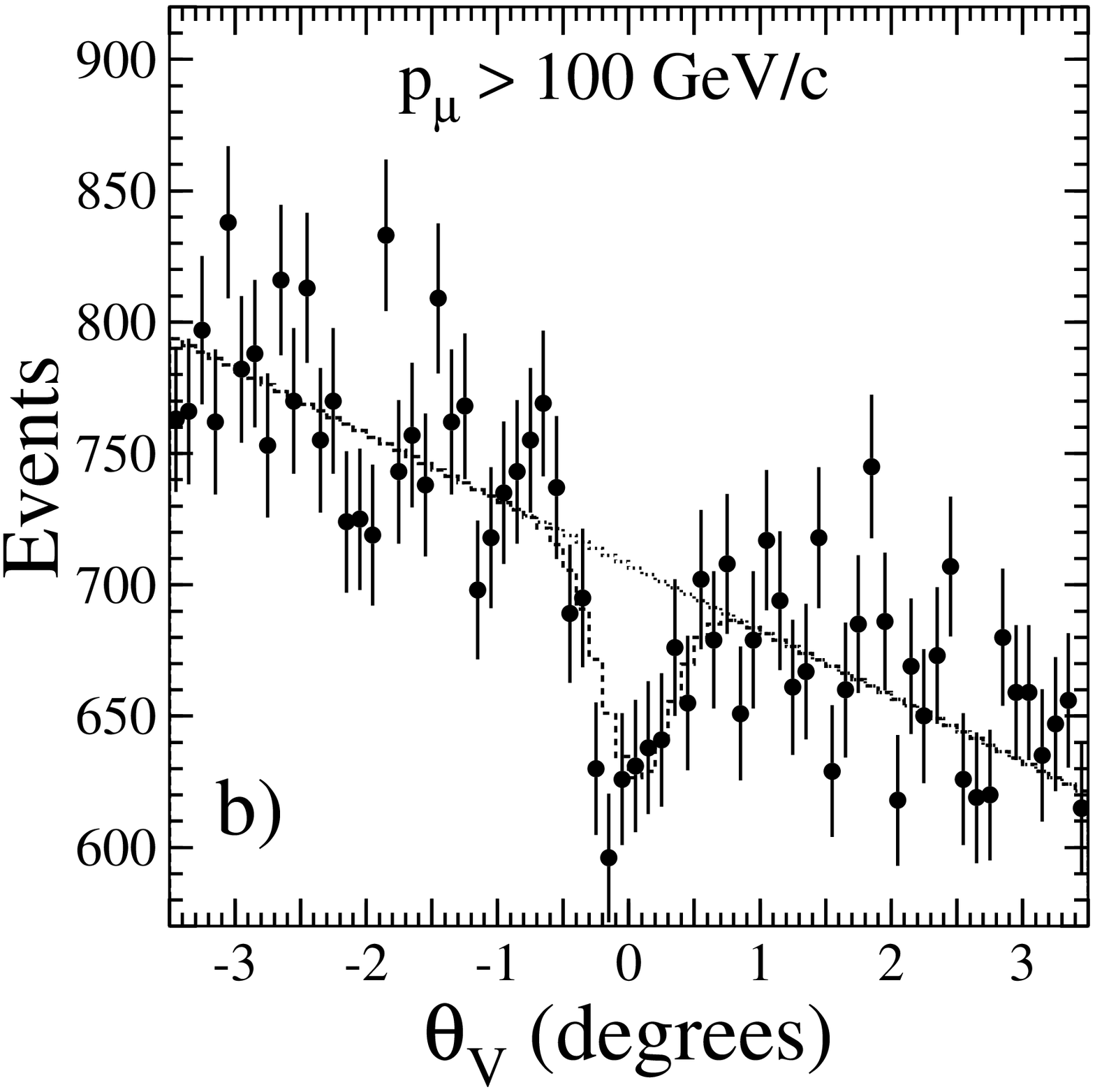} \\
\caption{ \footnotesize{Comparison between data and fitted results for 
the HE sample in a projection on a) the horizontal axis, b) the vertical axis.
Differences in the projected bandwidths (a: 0.8 degrees, b: 3.0 degrees)
explain the differences of the observed
number of events. The dotted lines represent the fitted average muon event number outside of the Moon region,
and the dashed lines the fit to the data in the Moon region.
}}
\label{pro_data_HE}
\end{center}
\end{figure}

Values of $N_{\rm miss}$ are extracted from the fit.
However $N_{\rm miss}$ can be also directly deduced from $\Phi$,
the flux measurement around the Moon direction,
 $N_{\rm miss}= \Phi \times T_{\rm live} \times \Omega_{\rm Moon}$
where $T_{\rm live}$ is the live-time corresponding to the Moon observation.
The main contribution to the uncertainty is related to the Moon solid
angle. At any time the precision of the calculations of the Moon radius $R_{\rm M}$, from
the SLALIB subroutines \cite{slalib}, is estimated to be 0.4\,\% and 
the time distribution of the Moon events is very well known. The uncertainty on the solid angle,
proportional to $R_{\rm M}^{2}$ is thus 0.8\,\%. The flux is known at the level of
0.3\,\% and the uncertainty on the live-time is still smaller. In total,
the precision on the most probable value of $N_{\rm miss}$ is estimated to be 
around 1\,\%. 
The expected values of $N_{\rm miss}$ are shown in Table \ref{result_7par}.
This knowledge of $N_{\rm miss}$ is introduced in the likelihood function
and allows an improvement in the determination of the remaining parameter $\sigma$.
$N_{\rm miss}$ is constrained to its most probable value, with a 1\,\% uncertainty.
The results are illustrated in Figure \ref{likevssigma}b for the LE sample
with the plot of $\Delta \ln\mathcal{L}$ versus $\sigma$. 
The experimental results 
are  $\sigma = (0.22 \, \pm  0.04)^{\circ}$ for the HE sample and
$\sigma = (0.28\ ^{+0.08}_{-0.05})^{\circ}$ for the LE sample.
These numbers refer to an effective angular resolution valid for
the set of selected events, integrated over the momentum distribution of the
data and the directional range of the Moon events.
For the HE sample, Figure \ref{pro_data_HE} shows 
a comparison of a projected band of data around the nominal position of
the Moon with the fitted results corresponding to
Table \ref{result_7par}.

As a cross check, the effective angular resolution is also obtained by a fit to
the one-dimensional deficit distribution. The values discussed above are confirmed,
albeit with much larger uncertainties.

\paragraph{The antiproton search }

To set a limit on a possible cosmic ray antiproton component,
the number of missing events $N_{\rm miss}$ is supposed to be shared between protons,
helium and antiprotons.
The total number of missing events has been constrained to the
expected value. The antiproton deficit is
described like the proton deficit with the corresponding parametrisation 
function symmetric to the proton one with respect to the Moon position.

As mentioned in section \ref{sec-sg-mu}, the value of $\sigma$ can be deduced from the
simulation. Comparison with data in di-muon events shows a good
agreement. Contrary to the situation in the
Moon-shadow experiment, high statistics is available and detailed
investigations are possible.
The obtained results are $\sigma = (0.24 \pm 0.01)^{\circ}$ for the HE sample and
$\sigma = (0.38 \pm 0.02)^{\circ}$ for the LE sample.
Uncertainties are better than those obtained above.
This angular information is
used and a maximum likelihood fit is performed using the $\rm \Pap$ content as a free parameter.

\begin{figure}[!ht]
\begin{center}
\includegraphics[height=12cm]{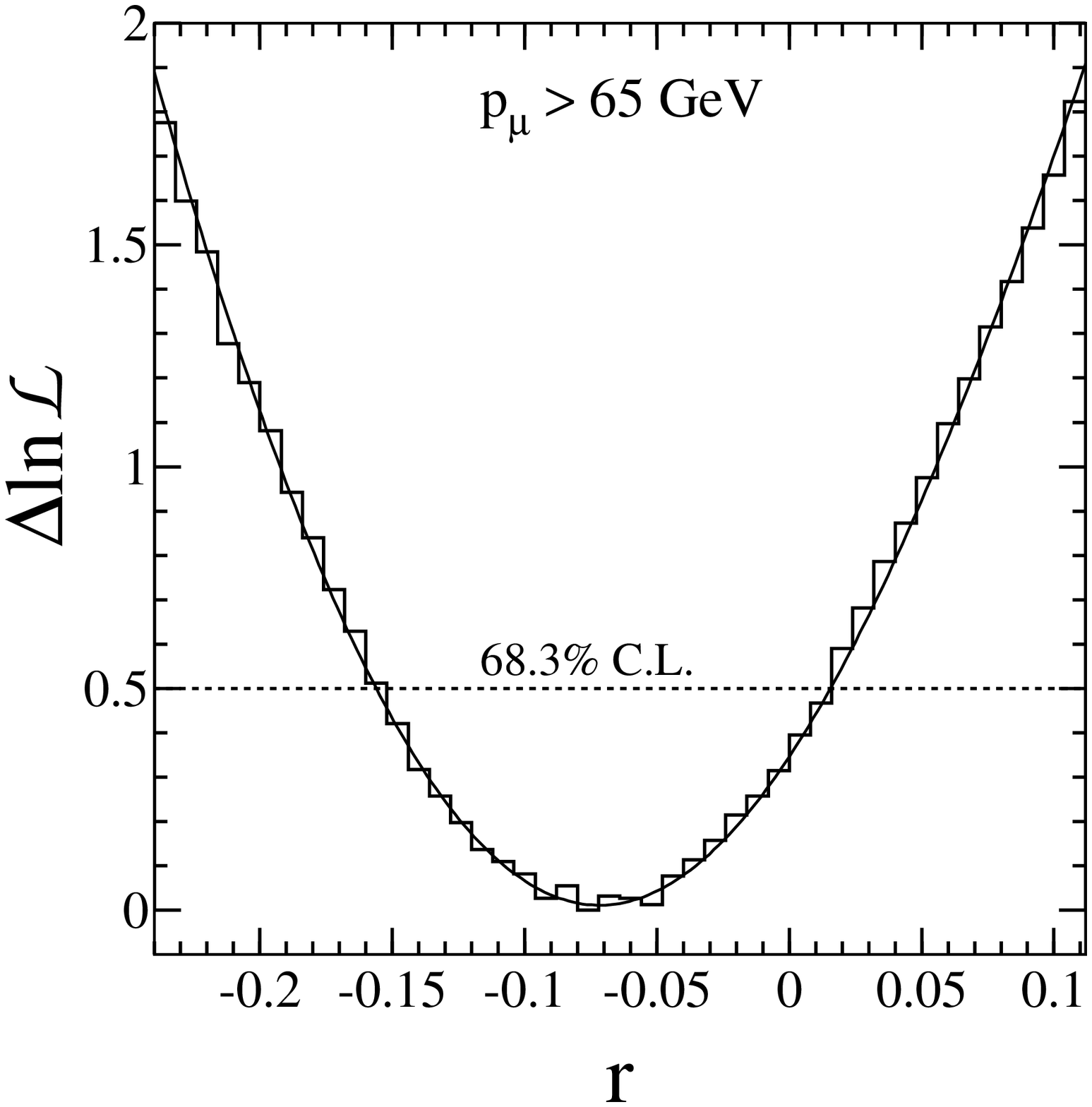}
\caption{ \footnotesize{$\Delta \ln\mathcal{L}$ as a function of r (the $\rm \Pap$ 
content) for the whole data.
The dashed line $\Delta \ln\mathcal{L} = 0.5$
is used to determine the $68.3\%$ central confidence interval.
}}
\label{likevsr_tot}
\end{center}
\end{figure}

\begin{figure}[!ht]
\begin{center}
\includegraphics[bb= 1 1 504 517,clip,width=15 cm,height=15cm]{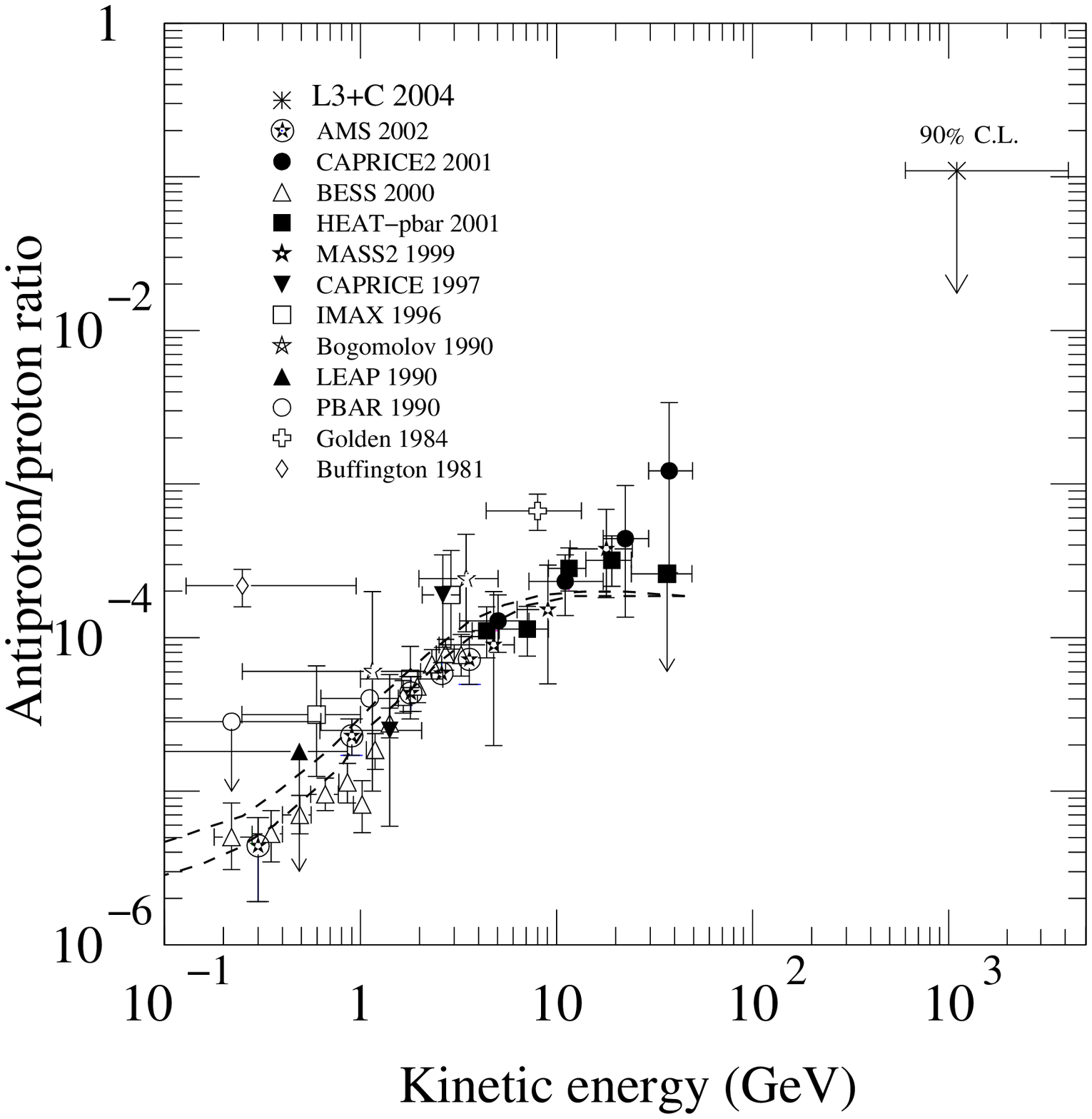}
\caption{\footnotesize{Measurements of the ratio of the antiproton and proton fluxes versus 
the primary energy, including the L3+C limit around 1 TeV. (The dashed lines show the range
of the theoretical expectations\,\protect\cite{moska}.)}}
\label{result}
\end{center}
\end{figure}

The HE and LE results are combined to give the final
measurement. This is done simply by adding the likelihood 
logarithmic-functions of each range.
The total significance of the deficit is $9.4~ s.d$.
The uncertainty range is obtained by 
finding the points for which $\Delta \ln\mathcal{L}\,=\,\Delta 
\ln\mathcal{L}_{\rm m}+0.5$. The result is shown in Figure \ref{likevsr_tot} and one finds
$r \, = \, \phi_{\Pap}/\phi_{\rm matter} \, = \,-0.07\, \pm \, 0.09 $. 
The result is below a physical boundary
(the $\rm \Pap$ content must be positive). An upper limit of 0.08 with 90\,\%  confidence level
is set using the unified approach \cite{feldman1}.
With the assumed flux composition around 1 TeV of 75\,\% protons and 25\,\% heavier nuclei
responsible of the observed deficit,
this corresponds to a $\rm \overline{p}/p$ ratio of $r_{\rm \overline{p}/p} = 0.11$.
Figure \ref{result} shows the L3+C result together with other published  
values. 

%
%
%

\section{Conclusions}

The L3 detector has collected more than $10^{10}$ triggers of cosmic ray muons
during the years 1999 and 2000 in parallel with high energy particle physics
studies at the LEP accelerator at CERN. About
$6.7 \times 10^5$ events, with a direction pointing to a $5.0^{\circ}$ cone around the
Moon, are used and a Moon-shadow effect in cosmic rays is observed.

A two dimensional analysis confirms that
the alignment of the detector is correct to better than $0.2^{\circ}$ and that
the size and the shape of the deficit are compatible with the expectations.
Two sets of data corresponding to high ($E_{\mu}>100\rm\;GeV$)
and low energy muons ($\rm 65\;GeV<E_{\mu}<100\;GeV$) lead to values of the effective angular resolution
respectively of $\rm (0.22 \pm 0.04)^{\circ} $ and $\rm (0.28 \ ^{+0.08}_{-0.05})^{\circ}$.
These numbers include all effects due to the showering of the primary
cosmic ray in the atmosphere,
the multiple scattering in the molasse and the detector resolution.
They describe correctly the observed event deficit.
The observed significance of the Moon shadow effect is $9.4~ s.d$.
A significant effect due to the Earth magnetic field is
observed. This is better seen in a coordinate system with axis respectively
parallel and orthogonal to the deflection defined for each direction in the
local sky. The offset and the extension of the shadow are clearly dependent on the
muon momentum range considered.
With the hypothesis that the presence of antiprotons in cosmic rays would lead to 
a symmetric shadow to the one due to protons, a measurement
of the $\rm \Pap$ content is extracted from the data and is found to be $ r \, = \, -0.07 \pm 0.09 $.
A 90\,\% confidence level of $ 0.08 $ is set on $r$, corresponding to an antiproton over proton ratio of
$ r_{\rm \overline{p}/p} \, = \, 0.11$.

%
%
%
\section*{Acknowledgements}

We wish to 
acknowledge the contributions of all the engineers 
and technicians who have participated in the construction 
and maintenance of the L3 and L3+C experiments. 
Those of us who are not from member states
thank CERN for its hospitality and help. 

%
%
%
%
\bibliographystyle{/l3/paper/biblio/l3stylem}
\bibliography{%
/l3/paper/biblio/l3pubs,%
/l3/paper/biblio/aleph,%
/l3/paper/biblio/delphi,%
/l3/paper/biblio/opal,%
/l3/paper/biblio/markii,%
/l3/paper/biblio/otherstuff,%
l3paper}

\begin{thebibliography}{5}

\bibitem{clark}
G.W. Clark, {\sl Phys. Rev.} {\bf 108} (1957) 450.
%
\bibitem{cygnus1}
CYGNUS Collaboration, D.E. Alexandreas \etal, {\sl Phys. Rev.} {\bf D 43} (1991) 1735.
%
\bibitem{casa1}
CASA Collaboration, A. Borione \etal, {\sl Phys. Rev.} {\bf D 49} (1994) 1171. 
%
\bibitem{tibet1}
TIBET Collaboration, M. Amenomori \etal, {\sl Phys. Rev.} {\bf D 47} (1993) 2675.
%
\bibitem{macro1}
MACRO Collaboration, M. Ambrosio \etal, {\sl Phys. Rev.} {\bf D 59} (1999) 012003.
%
\bibitem{macro2}
MACRO Collaboration, M. Ambrosio \etal, {\sl Astropart. Phys.} {\bf20} (2003) 145.
%
\bibitem{soudan1}
SOUDAN 2 Collaboration, J.H. Cobb \etal, {\sl Phys. Rev.} {\bf D 61} (2000) 92002.
%
\bibitem{urban}
M. Urban \etal, {\sl Nucl. Phys. B (Proc. Suppl.)} {\bf 14B} (1990) 223.
%
\bibitem{moska}
I.V. Moskalenko, {\sl Astrophys. J.} {\bf 565} (2002) 280. 
%
\bibitem{caprice1}
CAPRICE Collaboration, M. Boezio \etal, {\sl Astrophys. J.} {\bf 561} (2001) 787. 
%
\bibitem{heat1}
HEAT Collaboration, A.S. Beach \etal, {\sl Phys. Rev. Lett.} {\bf 87} (2001) 271101.
%
%
\bibitem{ISVHECRI02}
L3+C collaboration, J.-F. Parriaud, in Proceedings of the XIVth Rencontres de Blois, 2002, Blois, France.\\ 
L3+C collaboration, P. Le Coultre, in Proceedings of the XIIth ISVHECRI, 2002, CERN, Geneva, Switzerland.\\
L3+C collaboration, P. Le Coultre, in Proceedings of the IVth NOW 2003, Kanasawa, Japan.\\
L3+C collaboration, Y.P. Xu, in HE 3.3.1., Proceedings of the XXVIIIth ICRC, 2003 Tsukuba, Japan.
\bibitem{TibetRatio}
TibetAS$\gamma$-collaboration, T. Kido, in HE 3.3.2., Proceedings of the XXVIIIth ICRC, 2003 Tsukuba, Japan.
%
\bibitem{steph}
S.A. Stephens, {\sl Astron. and Astrophys.} {\bf 149} (1985) 1.
%
\bibitem{ECRS}
L3+C collaboration, M. Unger, in Proceedings of the IXXth ECRS 2004, Florence, Italy. 
%
\bibitem{L3+Cspec}
L3 Collaboration, P. Achard \etal, {\sl Phys. Lett.} {\bf B 598} (2004) 15
%
\bibitem{sala}
A. Barrau \etal, {\sl Astron. and Astroph.} {\bf 388} (2002) 676.
%
\bibitem{ullio1}
L. Bergstr\H{o}m, J. Edsj\H{o} and P. Ullio, {\sl Astrophys. J.} {\bf 526} (1999) 215.
%
\bibitem{ullio2}
P. Ullio, Preprint, {\sl astro-ph/9904086} (1999).
%
\bibitem{stecker}
F.W. Stecker and A.W. Wolfendale, Proceedings of the 19th ICRC, La Jolla, USA (1985).
%
\bibitem{artemis}
D. Pomar\`ede \etal, {\sl Astropart. Phys.} {\bf 14} (2001) 287.
%
\bibitem{milagro_1}
MILAGRO Collaboration, R. Atkins \etal, {\sl Nucl. Instr. and Meth.} {\bf A 449} (2000) 478.
%
\bibitem{milagro_2}
M.O. Wascko, {\small PhD thesis, University of California, Riverside (2001)}.
%
\bibitem{milagro_pre}
F. Samuelson, Proceedings of the 27th ICRC, Hamburg, Germany (2001).
\bibitem{l3}
L3 Collaboration, B. Adeva \etal, {\sl Nucl. Instr. and Meth.} {\bf A 289} (1990) 35.
%
\bibitem{nim}
L3+C Collaboration, O. Adriani \etal, {\sl Nucl. Instr. and Meth.} {\bf A 488} (2002) 209.
%
\bibitem{Moonshadow} 
J.-F. Parriaud, PhD thesis, Universit\'e Claude Bernard - Lyon I, Lyon (2003).
%
\bibitem{geant}
GEANT Version 3.15; R. Brun \etal, preprint DD/EE/84-1 (1984), revised 1987.
%
\bibitem{corsika}
CORSIKA, Version 6.019; D. Heck \etal, Report FZKA 6019 (1998).
%
\bibitem{IGRF}
IGRF: International  Association of Geomagnetism and Astronomy (IAGA);
DGRF95.DAT and IGRF00.DAT for 1999 data, and IGRF00.DAT and IGRF00S.DAT for 2000 data.
%
\bibitem{wiebel}
B. Wiebel-Sooth \etal, {\sl Astron. and Astroph.} {\bf 330} (1998) 389.
%
\bibitem{runjob}
RUNJOB Collaboration, A.V. Apanasenko \etal, {\sl Astropart. Phys.} {\bf 16} (2001) 13.
%
\bibitem{AMS}
AMS Collaboration, M. Aguilar \etal, {\sl Phys. Rep.} {\bf 366} (2002) 331; {\sl Phys. Lett.} {\bf B490} (2000) 27.
%
\bibitem{BESSc}
BESS collaboration, S.Haino \etal, {\sl Phys. Lett.} {\bf B594} (2004) 35.
%
\bibitem{slalib}
SLALIB - A positional Astronomy Library, Rutherford Appleton Laboratory,\\
$http://www.eso.org/science/scisoft/star/sun67.htx/sun67.html$.
%
\bibitem{feldman1}
G. Feldman and R. Cousins, {\sl Phys. Rev.} {\bf D 57} (1998) 3873.
%


\end{thebibliography}

%
%
\newpage

\typeout{   }     
\typeout{Using author list for paper 287 -  }
\typeout{$Modified: Jul 15 2001 by smele $}
\typeout{!!!!  This should only be used with document option a4p!!!!}
\typeout{   }
%
%
%
%
%
%

\newcount\tutecount  \tutecount=0
\def\tutenum#1{\global\advance\tutecount by 1 \xdef#1{\the\tutecount}}
\def\tute#1{$^{#1}$}
\tutenum\aachen            
\tutenum\nikhef            
\tutenum\mich              
\tutenum\lapp              
\tutenum\basel             
\tutenum\lsu               
\tutenum\beijing           
\tutenum\itp               
\tutenum\hum               
\tutenum\bologna           
\tutenum\tata              
\tutenum\ne                
\tutenum\bucharest         
\tutenum\budapest          
\tutenum\mit               
\tutenum\panjab            
\tutenum\debrecen          
\tutenum\dublin            
\tutenum\florence          
\tutenum\cern              
\tutenum\wl                
\tutenum\geneva            
\tutenum\hamburg           
\tutenum\hefei             
\tutenum\lausanne          
\tutenum\lyon              
\tutenum\madrid            
\tutenum\florida           
\tutenum\milan             
\tutenum\moscow            
\tutenum\naples            
\tutenum\cyprus            
\tutenum\nymegen           
\tutenum\osaka             
\tutenum\caltech           
\tutenum\perugia           
\tutenum\peters            
\tutenum\cmu               
\tutenum\potenza           
\tutenum\prince            
\tutenum\riverside         
\tutenum\rome              
\tutenum\salerno           
\tutenum\ucsd              
\tutenum\santiago          
\tutenum\sofia             
\tutenum\korea             
\tutenum\taiwan            
\tutenum\tsinghua          
\tutenum\purdue            
\tutenum\psinst            
\tutenum\zeuthen           
\tutenum\eth               

{
\parskip=0pt
\noindent
{\bf The L3 Collaboration:}
\ifx\selectfont\undefined
 \baselineskip=10.8pt
 \baselineskip\baselinestretch\baselineskip
 \normalbaselineskip\baselineskip
 \ixpt
\else
 \fontsize{9}{10.8pt}\selectfont
\fi
\medskip
\tolerance=10000
\hbadness=5000
\raggedright
\hsize=162truemm\hoffset=0mm
\def\r{\rlap,}
\noindent

P.Achard\r\tute\geneva\ 
O.Adriani\r\tute{\florence}\ 
M.Aguilar-Benitez\r\tute\madrid\
M.van~den~Akker\r\tute\nymegen\ 
J.Alcaraz\r\tute{\madrid}\ 
G.Alemanni\r\tute\lausanne\
J.Allaby\r\tute\cern\
A.Aloisio\r\tute\naples\ 
M.G.Alviggi\r\tute\naples\
H.Anderhub\r\tute\eth\
% ...

\newpage
\begin{list}{A}{\itemsep=0pt plus 0pt minus 0pt\parsep=0pt plus 0pt minus 0pt
                \topsep=0pt plus 0pt minus 0pt}
\item[\aachen]
 III. Physikalisches Institut, RWTH, D-52056 Aachen, Germany$^{\S}$
\item[\nikhef] NIKHEF, 
     and University of Amsterdam, NL-1009 DB Amsterdam, The Netherlands
\item[\mich] University of Michigan, Ann Arbor, MI 48109, USA
\item[\lapp] LAPP,IN2P3-CNRS, BP 110, F-74941 Annecy-le-Vieux CEDEX, France
\item[\basel] Institute of Physics, University of Basel, CH-4056 Basel,
     Switzerland
\item[\lsu] Louisiana State University, Baton Rouge, LA 70803, USA
\item[\beijing] Institute of High Energy Physics, IHEP, 
  100039 Beijing, China$^{\triangle}$ 
\item[$\itp$] ITP, Academia Sinica, 100039 Beijing, China 
\item[$\hum$] Humboldt University, D-10115 Berlin, Germany. 
\item[\bologna] University of Bologna and INFN-Sezione di Bologna, 
     I-40126 Bologna, Italy
\item[\tata] Tata Institute of Fundamental Research, Mumbai (Bombay) 400 005, India
\item[\ne] Northeastern University, Boston, MA 02115, USA
\item[\bucharest] Institute of Atomic Physics and University of Bucharest,
     R-76900 Bucharest, Romania
\item[\budapest] Central Research Institute for Physics of the 
     Hungarian Academy of Sciences, H-1525 Budapest 114, Hungary$^{\ddag}$
\item[\mit] Massachusetts Institute of Technology, Cambridge, MA 02139, USA
\item[\panjab] Panjab University, Chandigarh 160 014, India
\item[\debrecen] KLTE-ATOMKI, H-4010 Debrecen, Hungary$^\P$
\item[\dublin] Department of Experimental Physics,
  University College Dublin, Belfield, Dublin 4, Ireland
\item[\florence] University of Florence and INFN, Sezione di Firenze,  
     I-50019 Sesto Fiorentino, Italy
\item[\cern] European Laboratory for Particle Physics, CERN, 
     CH-1211 Geneva 23, Switzerland
\item[\wl] World Laboratory, FBLJA  Project, CH-1211 Geneva 23, Switzerland
\item[\geneva] University of Geneva, CH-1211 Geneva 4, Switzerland
\item[\hamburg] University of Hamburg, D-22761 Hamburg, Germany
\item[\hefei] Chinese University of Science and Technology, USTC,
      Hefei, Anhui 230 029, China$^{\triangle}$
\item[\lausanne] University of Lausanne, CH-1015 Lausanne, Switzerland
\item[\lyon] Institut de Physique Nucl\'eaire de Lyon, 
     IN2P3-CNRS,Universit\'e Claude Bernard, 
     F-69622 Villeurbanne, France
\item[\madrid] Centro de Investigaciones Energ{\'e}ticas, 
     Medioambientales y Tecnol\'ogicas, CIEMAT, E-28040 Madrid,
     Spain${\flat}$ 
\item[\florida] Florida Institute of Technology, Melbourne, FL 32901, USA
\item[\milan] INFN-Sezione di Milano, I-20133 Milan, Italy
\item[\moscow] Institute of Theoretical and Experimental Physics, ITEP, 
     Moscow, Russia
\item[\naples] INFN-Sezione di Napoli and University of Naples, 
     I-80125 Naples, Italy
\item[\cyprus] Department of Physics, University of Cyprus,
     Nicosia, Cyprus
\item[\nymegen] Radboud University and NIKHEF, 
     NL-6525 ED Nijmegen, The Netherlands
\item[$\osaka$] Osaka City University, Osaka 558-8585, Japan     
\item[\caltech] California Institute of Technology, Pasadena, CA 91125, USA
\item[\perugia] INFN-Sezione di Perugia and Universit\`a Degli 
     Studi di Perugia, I-06100 Perugia, Italy        
\item[\peters] Nuclear Physics Institute, St. Petersburg, Russia
\item[\cmu] Carnegie Mellon University, Pittsburgh, PA 15213, USA
\item[\potenza] INFN-Sezione di Napoli and University of Potenza, 
     I-85100 Potenza, Italy
\item[\prince] Princeton University, Princeton, NJ 08544, USA
\item[\riverside] University of California, Riverside, CA 92521, USA
\item[\rome] INFN-Sezione di Roma and University of Rome, ``La Sapienza",
     I-00185 Rome, Italy
\item[\salerno] University and INFN, Salerno, I-84100 Salerno, Italy
\item[\ucsd] University of California, San Diego, CA 92093, USA
\item[$\santiago$] University of Santiago de Compostela, E-15706 Santiago, Spain
\item[\sofia] Bulgarian Academy of Sciences, Central Lab.~of 
     Mechatronics and Instrumentation, BU-1113 Sofia, Bulgaria
\item[\korea]  The Center for High Energy Physics, 
     Kyungpook National University, 702-701 Taegu, Republic of Korea
\item[\taiwan] National Central University, Chung-Li, Taiwan, China
\item[\tsinghua] Department of Physics, National Tsing Hua University,
      Taiwan, China
\item[\purdue] Purdue University, West Lafayette, IN 47907, USA
\item[\psinst] Paul Scherrer Institut, PSI, CH-5232 Villigen, Switzerland
\item[\zeuthen] DESY, D-15738 Zeuthen, Germany
\item[\eth] Eidgen\"ossische Technische Hochschule, ETH Z\"urich,
     CH-8093 Z\"urich, Switzerland

\item[\S]  Supported by the German Bundesministerium 
        f\"ur Bildung, Wissenschaft, Forschung und Technologie.
\item[\ddag] Supported by the Hungarian OTKA fund under contract
numbers T019181, F023259 and T037350.
\item[\P] Also supported by the Hungarian OTKA fund under contract
  number T026178.
\item[$\flat$] Supported also by the Comisi\'on Interministerial de Ciencia y 
        Tecnolog{\'\i}a.
\item[$\sharp$] Also supported by CONICET and Universidad Nacional de La Plata,
        CC 67, 1900 La Plata, Argentina.
\item[$\triangle$] Supported by the National Natural Science
  Foundation of China.
\item[$\odot$] On leave from the Moscow Physical Engineering Institute (MePhl).
\item[$\diamond$] On leave from JINR, RU-141980 Dubna, Russia. 
\item[$\dagger$] Deceased.
\item[$\oplus$] Corresponding author, e-mail: $Pierre.Le.Coultre@cern.ch$
\end{list}
}
\vfill


\end{document}